\documentclass[aps, pra, twocolumn, showpacs,footinbib,superscriptaddress]{revtex4-1}
\usepackage{graphicx, subfigure,braket}
\usepackage{amsmath,amssymb}
\usepackage{color}
\usepackage{natbib,hyperref}

\usepackage{placeins}

\begin{document}

\title{Neural-network-designed three-qubit gates robust against charge noise and crosstalk in silicon}

\author{David W. Kanaar}
\affiliation{Department of Physics, University of Maryland Baltimore County, Baltimore, MD 21250, USA}
%\author{Utkan G\"ung\"ord\"u}
%\affiliation{Laboratory for Physical Sciences, College Park, Maryland 20740, USA}
%\affiliation{Department of Physics, University of Maryland, College Park, Maryland 20742, USA}
\author{J.~P.~Kestner}
\affiliation{Department of Physics, University of Maryland Baltimore County, Baltimore, MD 21250, USA}

\begin{abstract}
Spin qubits in semiconductor quantum dots are a promising platform for quantum computing, however scaling to large systems is hampered by crosstalk and charge noise. Crosstalk here refers to the unwanted off-resonant rotation of idle qubits during the resonant rotation of the target qubit. For a three-qubit system with crosstalk and charge noise, it is difficult to analytically create gate protocols that produce three-qubit gates, such as the Toffoli gate, directly in a single shot instead of through the composition of two-qubit gates. Therefore, we numerically optimize a physics-informed neural network to produce theoretically robust shaped pulses that generate a Toffoli-equivalent gate. Additionally, robust $\frac{\pi}{2}$ $X$ and {\sc cz} gates are also presented in this work to create a universal set of gates robust against charge noise. The robust pulses maintain an infidelity of $10^{-3}$ for average quasistatic fluctuations in the voltage of up to a few mV instead of tenths of mV for non-robust pulses.  %detail time length scale %detail same bandwitch but more driving strength
\end{abstract}

\maketitle

\section{Introduction}
Spin qubits in silicon are a promising platform for creating scalable quantum computers. Current single- and two-qubit gate infidelities have been reduced to the order of $10^{-3}$ or below \cite{Yang2019,Yoneda2018,Petit_2020,xue2021,Noiri2022,mills2021twoqubit,Huang2019}, which is under the surface code threshold \cite{Fowler2012} and is nearing the $10^{-4}$ threshold for CSS quantum error correcting codes \cite{NielsenandChuang}. Achieving similar fidelities in silicon devices with more than two qubits is challenging due to crosstalk and increased susceptibility to charge noise in these devices. Crosstalk here refers to the unwanted rotation of qubits other than the target qubit as a result of the difference in resonance frequencies between qubits not being sufficiently large compared to the driving strength (i.e. the maximum Rabi frequency). 

Universal quantum computation requires a universal set of gates, for instance, single-qubit rotations and a Toffoli gate \cite{Kitaev1997}.
Additionally, a Toffoli gate has a classical analog and is used in many circuits for classical operations such a binary arithmetic \cite{Shende2008,Maslov2016}. Having a direct implementation of a Toffoli gate would be faster than implementing Toffoli gates using a sequence of single- and two-qubit gates. It is, however, not straightforward to design a Toffoli gate for a chain of three exchange-coupled spins. It was shown that an equivalent gate, the iToffoli gate, can be implemented in exchange coupled spins using a pulse sequence in the absence of crosstalk \cite{Gullans_2019}. Unfortunately, for a scalable silicon qubit system, it is not always possible to ignore crosstalk \cite{seedhouse2021}. Methods for dealing with crosstalk for two-qubit gates have been proposed \cite{Hansen2021,Heinz2021}. However, these methods cannot be logically extended for a direct implementation of a Toffoli-equivalent gate.
Furthermore, the fidelity of gates in silicon devices is limited by charge noise \cite{huang_spin_2018,vanDijk2019} and previous methods do not correct for exchange fluctuations caused by charge noise. A method for correcting exchange fluctuations in \emph{two-qubit} systems while considering crosstalk has previously been proposed \cite{Gungordu2020}. However, this method cannot be logically extended to a three-qubit system.
Therefore, we present a shaped pulse which results in a Toffoli-equivalent gate using a physically informed neural network method originally presented in Ref.~\cite{Gungordu2020p2} and adapted to a two-qubit system in Ref.~\cite{Kanaar2022}. 

In Sec.~\ref{sec:Model}, the device Hamiltonian is shown and the neural network optimization method is described.  In Sec.~\ref{sec:resultinggates}, the pulse shapes for a non-robust iToffoli, a robust iToffoli, a robust Controlled-Z ({\sc cz}) gate, and a single-qubit $\frac{\pi}{2}$ $X$-rotation on the middle and outer qubits are presented. Either of the entangling gates plus the $\frac{\pi}{2}$ $X$-rotation, when combined with virtual-$Z$ rotations \cite{Mckay2017}, forms a universal set of gates robust against charge noise.

\section{Model}\label{sec:Model}
The device we consider is a chain of three exchange-coupled spins confined in silicon quantum dots. This device in the presence of magnetic fields is well described by the Heisenberg model Hamiltonian \cite{Loss1998} in Eq.~\eqref{eq:heis}. 
\begin{equation}
    H=\sum_{i=1}^{3} \mu_b g_{i} \mathbf{B}_{i} \cdot \mathbf{S}_{i}+ \sum_{i=1}^{2} J_{i} \left(\mathbf{S}_{i}\cdot \mathbf{S}_{i+1}-\frac{1}{4}\right) 
    \label{eq:heis}
\end{equation}
$\mu_b$ is the Bohr magneton, $g_i$ is the $g$-factor of the $i$-th qubit, $\mathbf{B}_i$ is the magnetic field at the $i$-th qubit, $\mathbf{S}_i$ is the spin operator of the $i$-th qubit,  $J_i$ is the exchange coupling between the $i$-th and $(i+1)$-th qubit,. The magnetic field is applied only in the $Z$ and $X$-directions and will can be described by the Zeeman energy  $E_{z,i}=\mu g_i B_{z,i}$ for the $Z$-direction and the envelope, $\Omega_i$, frequency, $\omega_i$, and phase $\phi_i$, $\Omega_i \cos(\omega_i t +\phi_i)=\mu g_i B_{x,i}$ for the $X$-direction.
The rotating frame Hamiltonian, $H_R=R H R^{\dagger}+i \hbar (\partial_t R) R^{\dagger}$, in the frame $R=\exp{\left(\frac{i}{2} \sum_{j=1}^3  (E_{z,j} t) Z_{j}\right)}$ becomes
\begin{widetext}
\begin{multline}\label{eq:Hdriven}
%H_R(t)= \frac{J_1(t)}{4} \left[Z_1 Z_2 + \cos ((E_{z,1}-E_{z,2}) t) \left(X_1 X_2+Y_1 Y_2\right) + \sin ((E_{z,1}-E_{z,2}) t) \left(X_1Y_2-Y_1X_2\right) \right]
%\\
%+ \frac{J_2(t)}{4} \left[Z_2 Z_3 + \cos ((E_{z,2}-E_{z,3}) t) \left(X_2 X_3+Y_2 Y_3\right) + \sin ((E_{z,2}-E_{z,3}) t) \left(X_2Y_3-Y_2X_3\right) \right]\\
%+ \frac{\Omega_1(t)}{2} [ \cos\phi_1(t) X_1 +\sin\phi_1(t) Y_1 +   \cos \left((E_{z,1}-E_{z,2}) t+\phi_1(t)\right) X_2 + \sin \left((E_{z,1}-E_{z,2}) t+\phi_1(t)\right) Y_2
%\\
%+   \cos \left((E_{z,1}-E_{z,3}) t+\phi_1(t)\right) X_3 + \sin \left((E_{z,1}-E_{z,3}) t+\phi_1(t)\right) Y_3 ]
%\\
%+ \frac{\Omega_2(t)}{2} [ \cos\phi_2(t) X_2 +\sin\phi_2(t) Y_2 +   \cos \left((E_{z,2}-E_{z,1}) t+\phi_2(t)\right) X_1 + \sin \left((E_{z,2}-E_{z,1}) t+\phi_2(t)\right) Y_1 
%\\
%+   \cos \left((E_{z,2}-E_{z,3}) t+\phi_2(t)\right) X_3 + \sin \left((E_{z,2}-E_{z,3}) t+\phi_2(t)\right) Y_3 ]
%\\
%+ \frac{\Omega_3(t)}{2} [ \cos\phi_3(t) X_3 +\sin\phi_3(t) Y_3 +   \cos \left((E_{z,3}-E_{z,1}) t+\phi_3(t)\right) X_1 + \sin \left((E_{z,3}-E_{z,1}) t+\phi_3(t)\right) Y_1 
%\\
%+   \cos \left((E_{z,3}-E_{z,2}) t+\phi_3(t)\right) X_2 + \sin \left((E_{z,3}-E_{z,2}) t+\phi_3(t)\right) Y_2 ] 
%\\
H_R(t)= \sum_{i=1}^3 \frac{\Omega_i}{2} \sum_{j=1}^{3} \cos((E_{z,i}-E_{z,j})t +\phi_{i}(t) )X_j+\sin((E_{z,i}-E_{z,j})t +\phi_{i}(t) )Y_j \quad 
\\
+ \sum_{i=1}^2 \frac{J_i(t)}{4} \left[ Z_{i}Z_{i+1} +\cos((E_{z,i}-E_{z,i+1})t)(X_{i}X_{i+1}+Y_{i}Y_{i+1})+\sin((E_{z,i}-E_{z,i+1})t)(X_{i}Y_{i+1}+Y_{i}X_{i+1}) \right],
\end{multline}
%\begin{equation}
%H_R(t)\approx \frac{\Omega_1}{2}(\cos(\phi_{1}(t) )X_1+\sin(\phi_{1}(t) )Y_1)+\frac{J_1(t)}{4} Z_{1}Z_{2}+\frac{J_2(t)}{4} Z_{2}Z_{3}
%\end{equation}
\end{widetext}
which is the three-qubit extension of the two-qubit Hamiltonian in Ref.~\cite{Kanaar2022}.
In Eq.~\eqref{eq:Hdriven} $X_i$, $Y_i$ and $Z_i$ are the Pauli operators on the $i$-th qubit with implied Kronecker products. Eq.~\eqref{eq:Hdriven} allows for three-tone driving, but in this work we will only use two-tone driving resonant with the outer qubits, i.e., $\Omega_2 = 0$. To demonstrate a pulse that can handle large crosstalk, the difference in Zeeman energies is chosen to be $E_{z,3}-E_{z,2}=E_{z,2}-E_{z,1}=5$MHz while the maximum driving strength is set to $\Omega_{\text{max}}=10$MHz such that they are of similar strength and the rotating wave approximation is certainly invalid. Currently, driving strengths of MHz have been implemented in experiment\cite{Yoneda2018,Watson2018,Croot2020,Noiri2022,Undseth2022,Gilbert2022}. The maximum exchange strength was chosen to be $J_{\text{max}}=10$MHz, which is readily attainable in experiments\cite{xue2021}. Additionally, the exchange will be taken to have a dependence on the barrier gate voltages $V_i$ as $J_i=J_0 e^{2 \alpha V_i}$ with $\alpha = 12.1 V^{-1}$ and $J_0=0.056$MHz as taken from Ref.~\cite{xue2021}. Our methods are not specific to these choices, but we simply wish to showcase the versatility of our numerical methods to incorporate this level of experimental detail.

\subsection{Neural network method}
The large number of non-commuting terms in this Hamiltonian \eqref{eq:Hdriven} make it very difficult to analytically create a pulse which results in a Toffoli or Toffoli-equivalent gate. Therefore, a pulse is found by numerically optimizing a physics-informed neural network using the method from Refs.~\cite{Gungordu2020p2,Kanaar2022}. The sole input of this neural network is time and the outputs are the control fields $J_i(t)$, $\Omega_i(t)$, and $\phi_i(t)$. These smooth control fields are optimized to result in gates robust against charge noise in the presence of crosstalk.
The DiffEqFlux.jl Julia package\cite{rackauckas2019diffeqfluxjl} was used to implement the neural network, the OrdinaryDiffEq.jl package \cite{Rackauckas2017} was used to implement the BS5 solver to solve the Schrodinger equation, and the Flux.jl and Optmin.jl packages were used to implement the RADAM and BFGS optimizers to minimize the cost function consisting of two terms. 

The first term in the cost function is the trace infidelity modulo virtual-$Z$ rotations \cite{Mckay2017},
\begin{equation}\label{eq:trace}
    1-F = \min_{\vec{\varphi}} \left[1-\frac{1}{8}\left|\text{Tr}\left(e^{\frac{i}{2}(\sum_{i=1}^{3} \varphi_i Z_i)} U_c e^{\frac{i}{2} (\sum_{i=4}^{6} \varphi_{i} Z_i)} U_{\text{t}}^{\dagger}\right)\right|^2\right],
\end{equation}
where $U_{\text{t}}$ is the targeted evolution operator, $U_{c}$ is the evolution operator produced by the control in the absence of noise, and $\varphi_i$ are the virtual-$Z$ rotation angles. 

The second term in the cost function quantifies the sensitivity of the evolution operator to noise sources. For instance, a given noise will couple to the qubits via a stochastic term in the Hamiltonian, $H_{\epsilon}$. The first-order Magnus expansion of the resulting noisy evolution operator is 
\begin{equation}\label{eq:MagnusU}
    U \approx U_c e^{\frac{i}{\hbar}\mathcal{E}}
\end{equation} 
where 
\begin{equation}\label{eq:MagnusE}
    \mathcal{E}= \int_0^T U_c^{\dagger} H_{\epsilon} U_c dt.
\end{equation}
Thus, we minimize sensitivity to that noise source by including the Frobenius norm of $\mathcal{E}$ in the cost function.

Charge noise effectively acts as fluctuations in the device voltages, which causes noise in the exchange couplings. The fluctuations in the two exchange couplings, $J_1$ and $J_2$, are assumed to be completely uncorrelated, as this is the worst case scenario for this type of optimization \cite{Jelmer2020}. This means the second term in the cost function contains the norm of the first-order Magnus expansions of both errors,
\begin{equation}\label{eq:cost}
    \mathcal{C} = \left(1-F\right) + w\times(|\mathcal{E}_1| + |\mathcal{E}_2|)
\end{equation}
where $w$ is a relative weighting factor which we set to be 0.1.
Minimizing only the first term in the cost function results in a gate that accounts for crosstalk in the absence of charge noise, while minimizing both terms adds robustness to charge noise. Using this method, a robust $\frac{\pi}{2}$ $X$ rotation on the outer and middle qubit, a robust $CZ$ gate, a non-robust iToffoli gate as well as a robust iToffoli gate are found. The iToffoli gate, $iT$, is a Toffoli gate that also adds an $i$ phase for the state (or part of the state) for which the controls are met, i.e., $iT\ket{a,b,c}\rightarrow i^{ab} \ket{a,b,c\otimes ab}$  where $a,b,c\in {0,1}$. An iToffoli gate with controls on the outside qubits was chosen because it was shown in Ref.~\cite{Gullans_2019} to be a more natural gate for this Hamiltonian than a Toffoli gate. The driving for all cases is chosen to be two-tone driving resonant with the outer two of the three qubits.

\section{Resulting  gates}\label{sec:resultinggates}
\subsection{$\frac{\pi}{2}$ $X$ gates}
\begin{figure}[t]
    \centering
    \includegraphics[width=\linewidth]{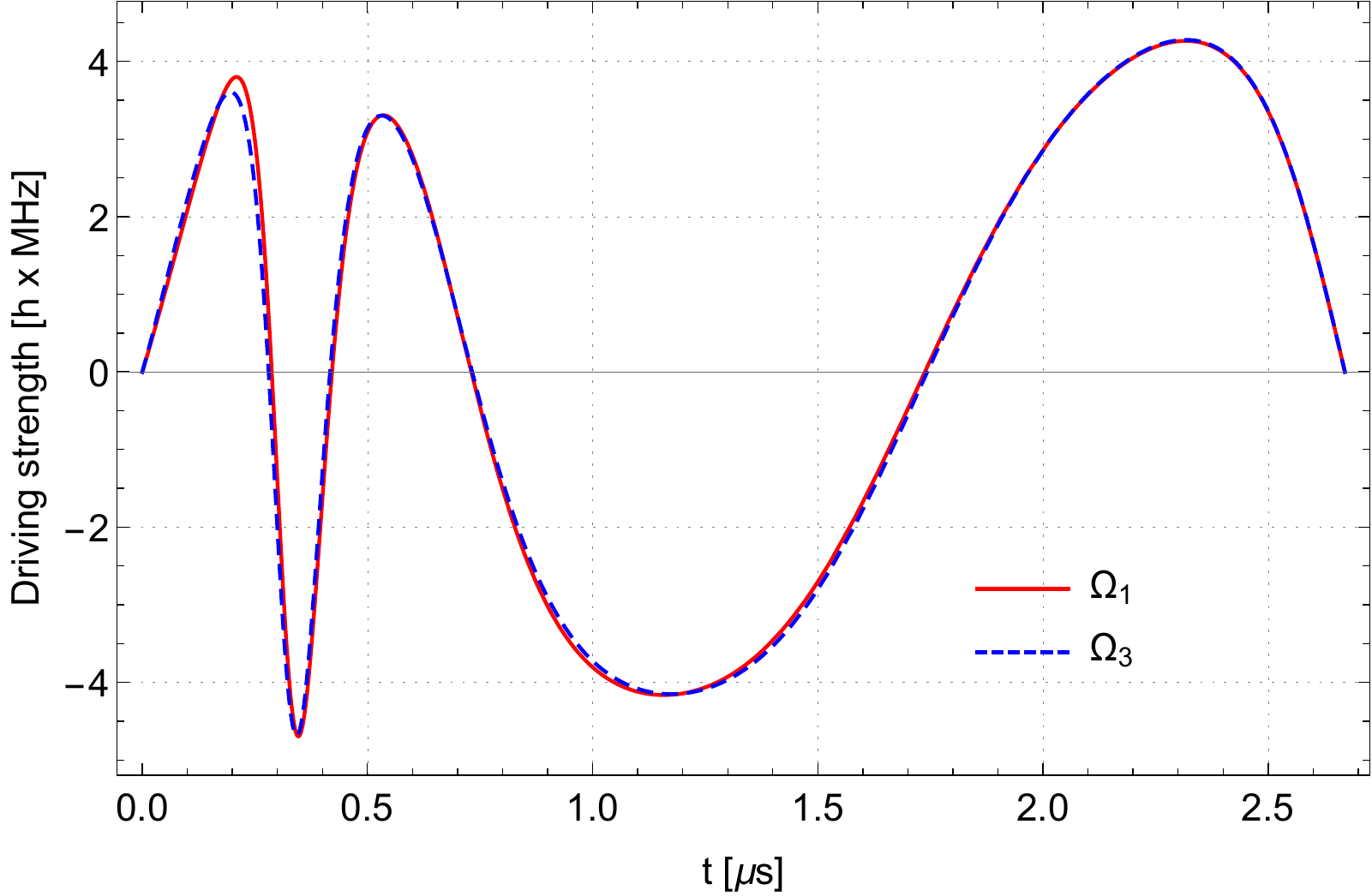}
    \includegraphics[width=\linewidth]{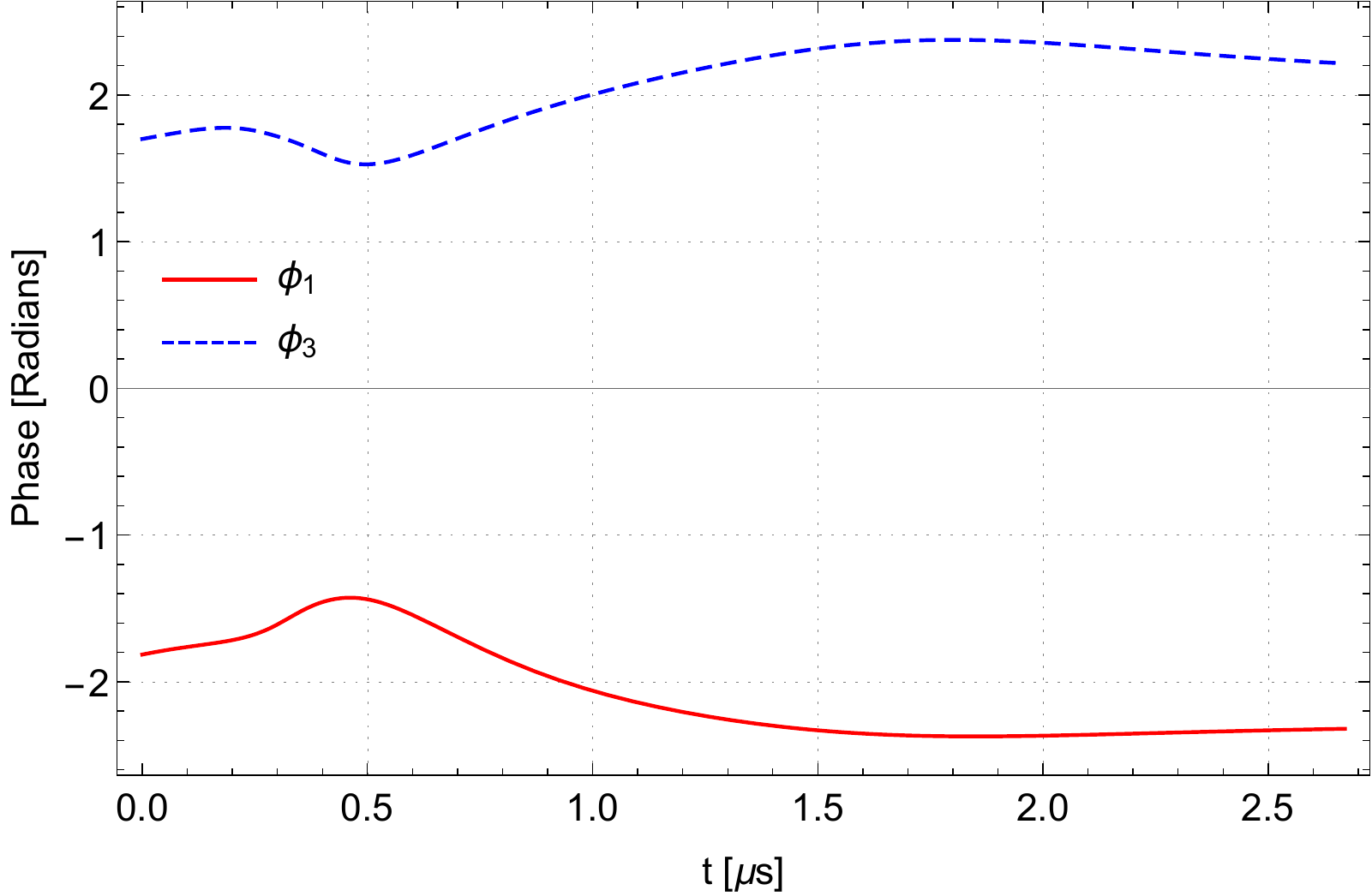}
      \caption{Neural network designed control fields vs time for a $\frac{\pi}{2}$ $X$ rotation on the middle qubit in the presence of crosstalk. Top: Driving tone amplitudes $\Omega_1$ and $\Omega_3$. Bottom: Accompanying phase modulation of the driving $\phi_1$ and $\phi_3$.}
      \label{fig:midX}
\end{figure}

\begin{figure}[t]
    \centering
    \includegraphics[width=\linewidth]{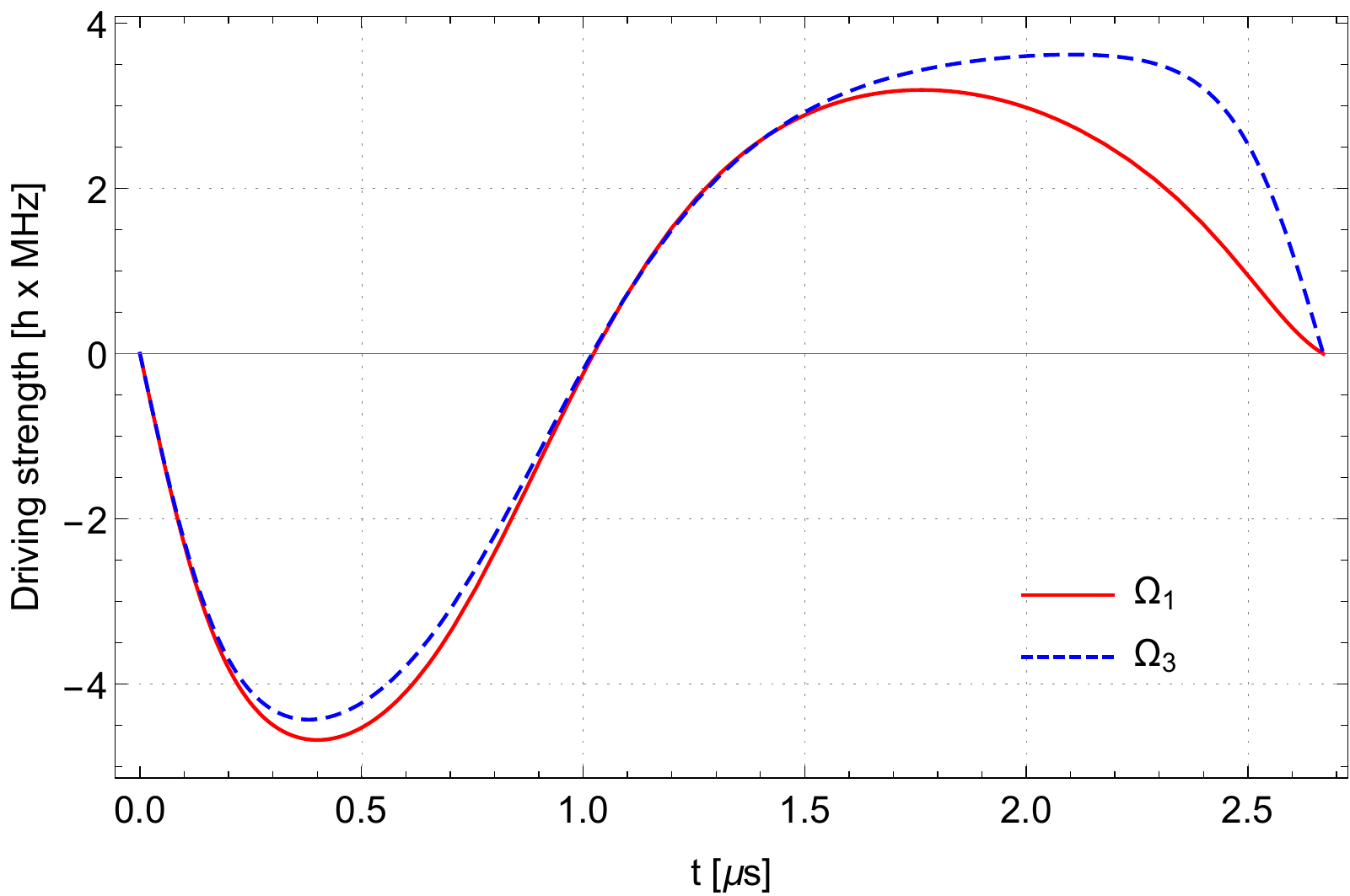}
    \includegraphics[width=\linewidth]{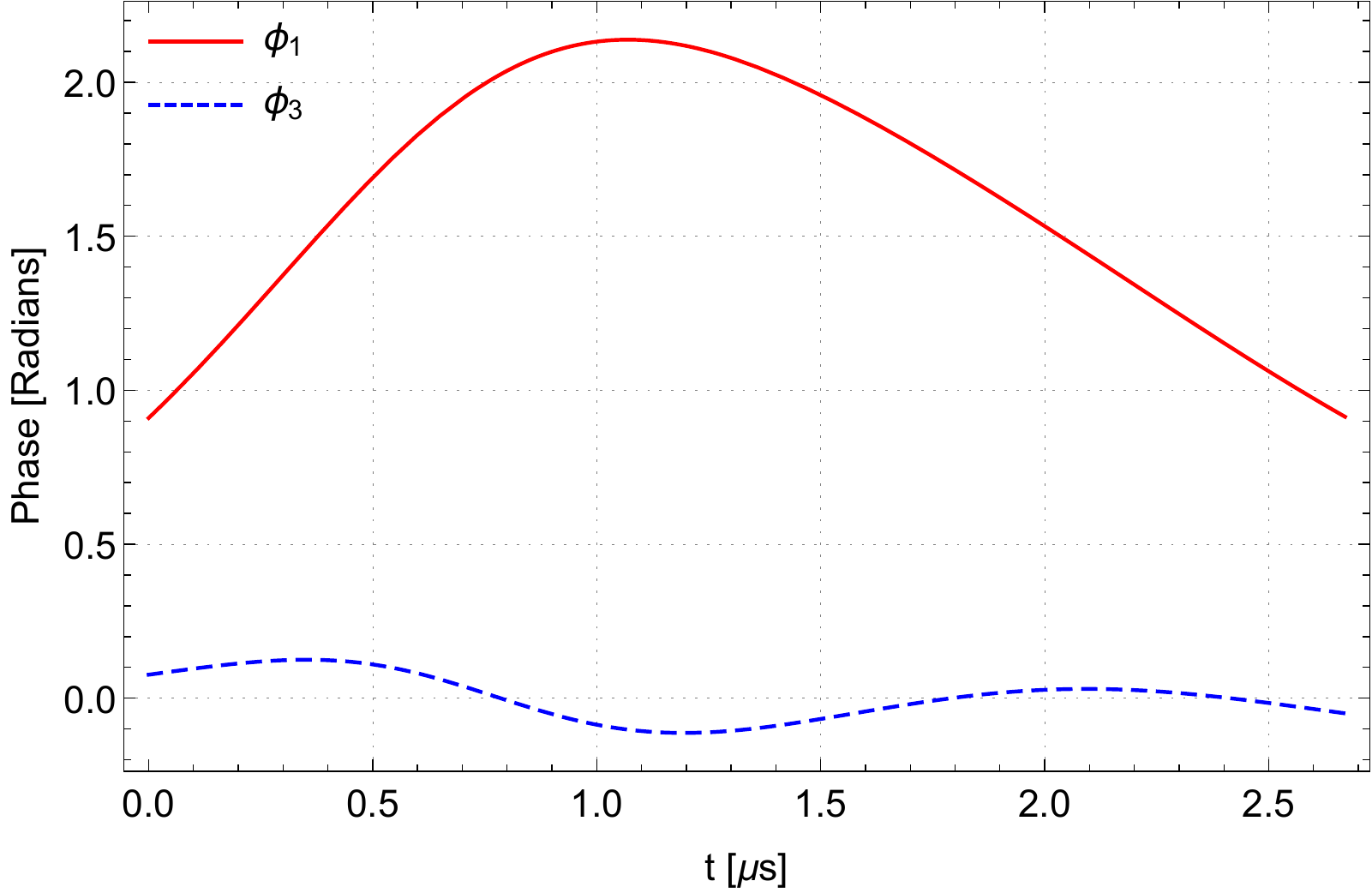}
      \caption{Neural network designed control fields vs time for a $\frac{\pi}{2}$ $X$ rotation on the third qubit in the presence of crosstalk. Top: Driving tone amplitudes $\Omega_1$ and $\Omega_3$. Bottom: Accompanying phase modulation of the driving tones $\phi_1$ and $\phi_3$.}
      \label{fig:outX}
\end{figure}

Local rotations do not require any entangling, therefore the control field pulses were optimized under the constraint that the exchange be held fixed as close to zero as possible. There is always some residual coupling -- here we take $J_1=J_2=J_0=0.056$MHz \cite{xue2021} -- but this is easily compensated for in the numerical optimization. Since the exchange is insensitive to voltage fluctuations around the uncoupled limit, any such local rotations are already trivially robust to charge noise. The optimized control field shapes for $\frac{\pi}{2}$ $X$ rotations on the middle and outer qubits are shown in Figs.~\ref{fig:midX} and \ref{fig:outX}, respectively. They maintain an infidelity below $10^{-4}$ even for unrealistically large quasistatic average barrier gate fluctuations, $(\delta V_1+\delta V_2)/2$, of 75mV (10mV) for the middle (outer) $X$ rotation.

\subsection{{\sc cz} gate}
\begin{figure}[t]
    \centering
    \includegraphics[width=\linewidth]{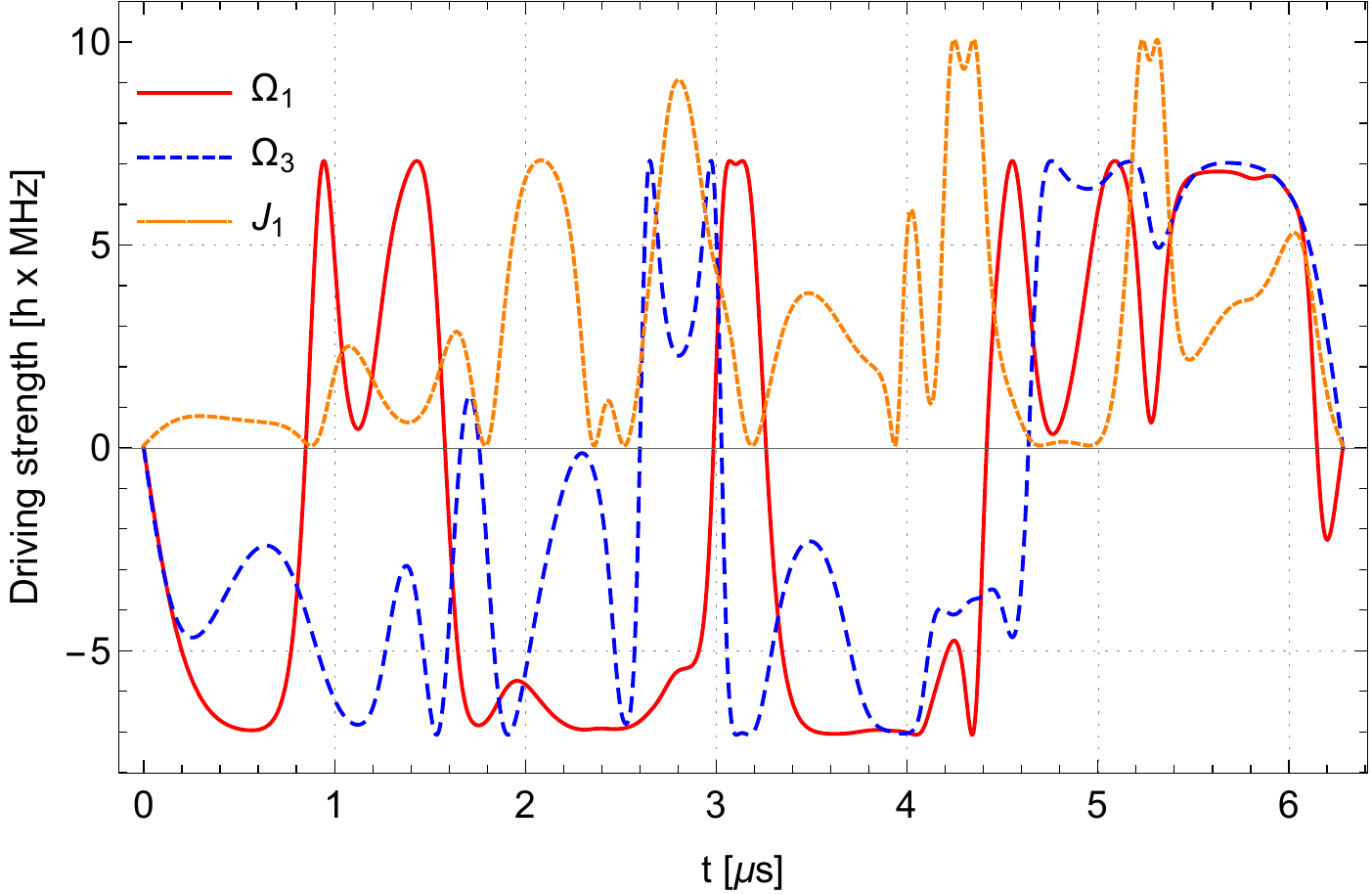}
    \includegraphics[width=\linewidth]{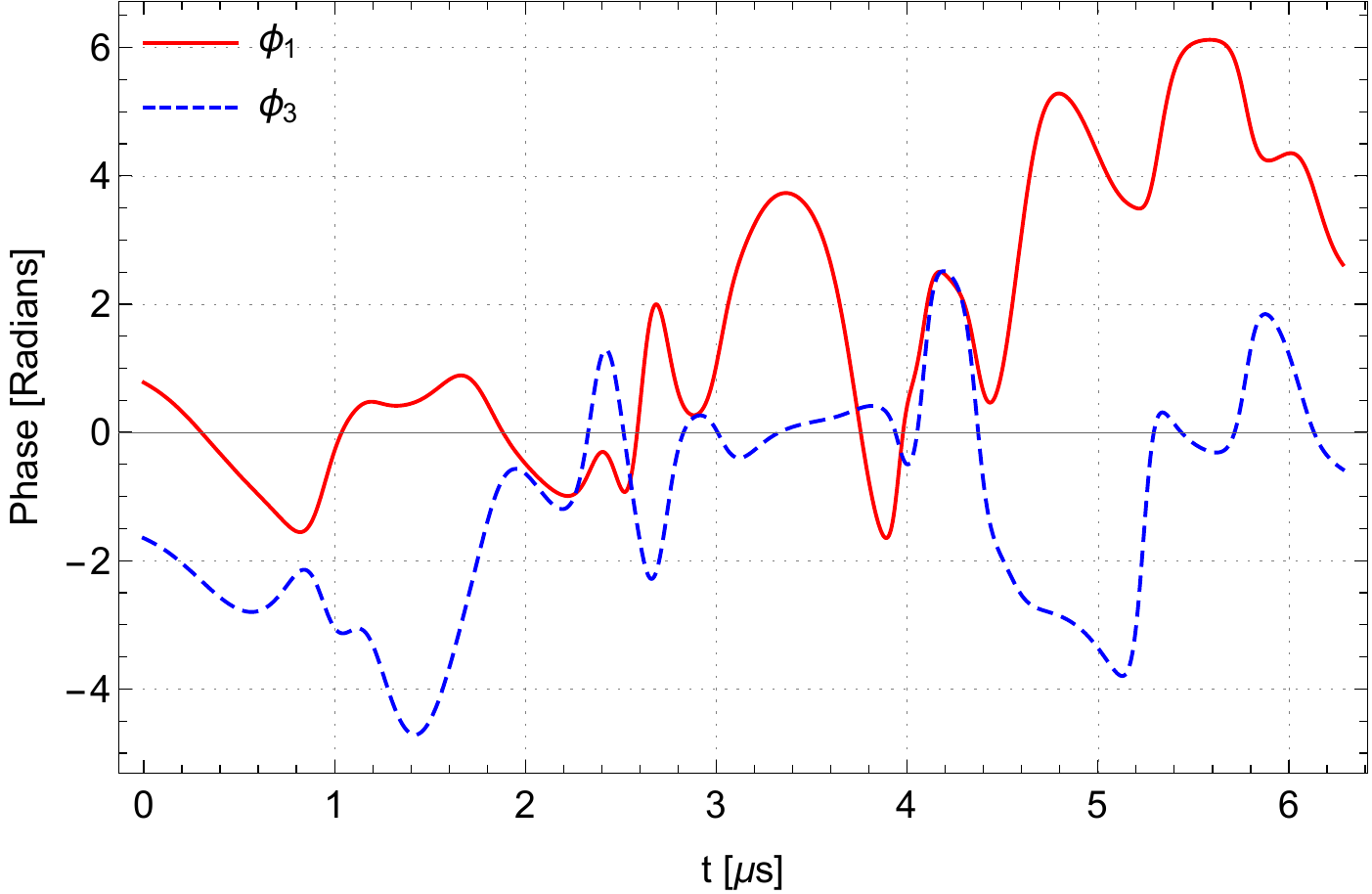}
      \caption{Neural network designed control fields vs time for a {\sc cz$_{12}$} gate in the presence of crosstalk. Top: Exchange amplitudes $J_1$ and $J_2$ as well as driving tone amplitudes $\Omega_1$ and $\Omega_3$. Bottom: Accompanying phase modulation of the driving tones $\phi_1$ and $\phi_3$.}
      \label{fig:ZZrob}
\end{figure}

A non-robust {\sc cz} gate is efficiently implemented adiabatically because nonadiabatic two-qubit gate fidelities are limited by the ratio of the difference in resonance frequencies to the exchange coupling strength used to implement the gate\cite{meunier_efficient_2011}. For this reason we only need to optimize for a robust {\sc cz} gate. The {\sc cz} gate on the first two qubits does not need exchange between the second and third qubit to be turned on, so we fix $J_2=J_0$. Therefore, the gate will automatically be insensitive to fluctuations in $\delta V_2$. The control fields of the robust {\sc cz} gate are shown in Fig.~\ref{fig:ZZrob}. 

\subsubsection{Charge noise analysis} \label{sec:chargenoiseanalysis1}
\begin{figure}[t]
    \centering
    \includegraphics[width=\linewidth]{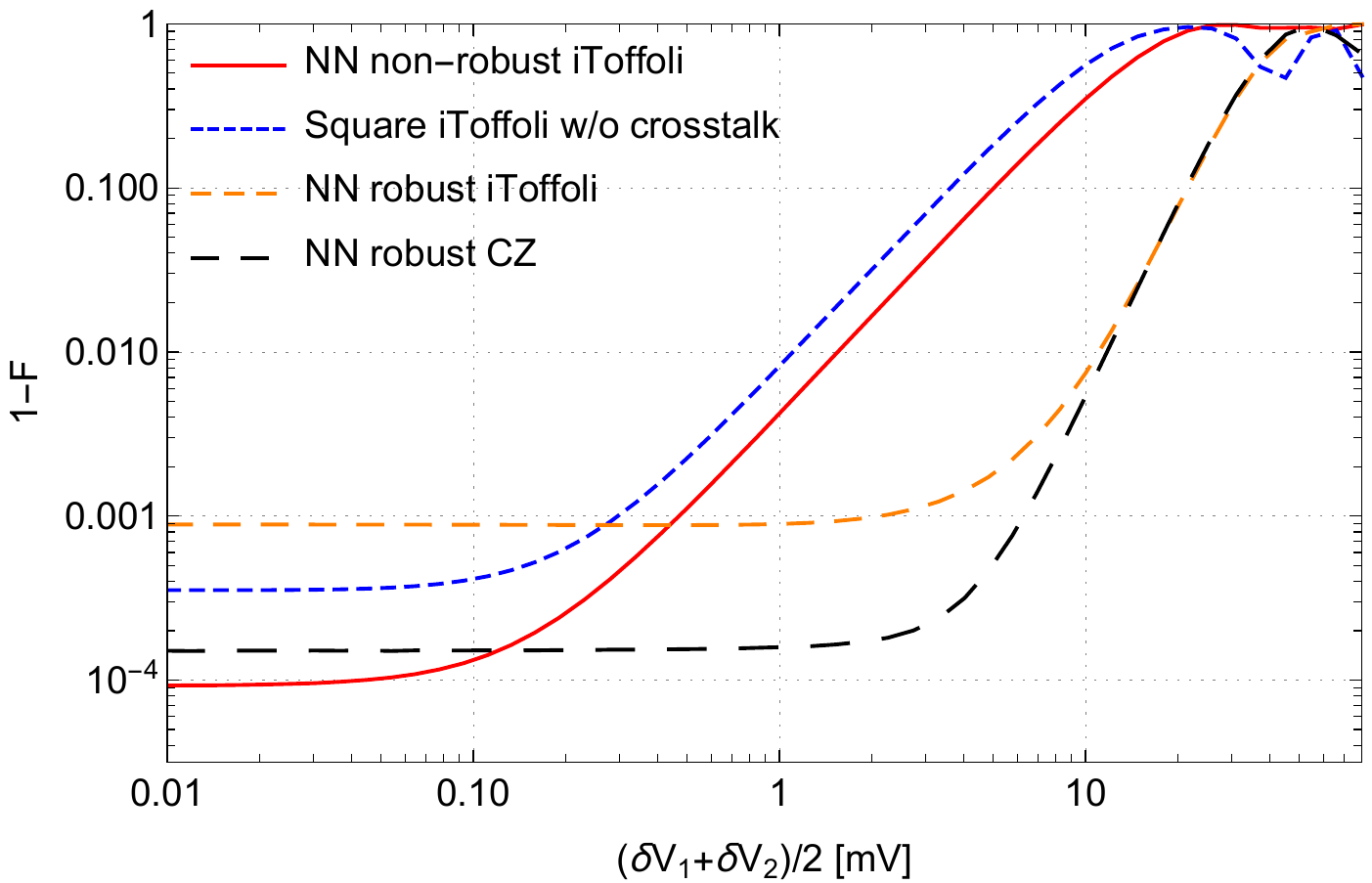}
    \includegraphics[width=\linewidth]{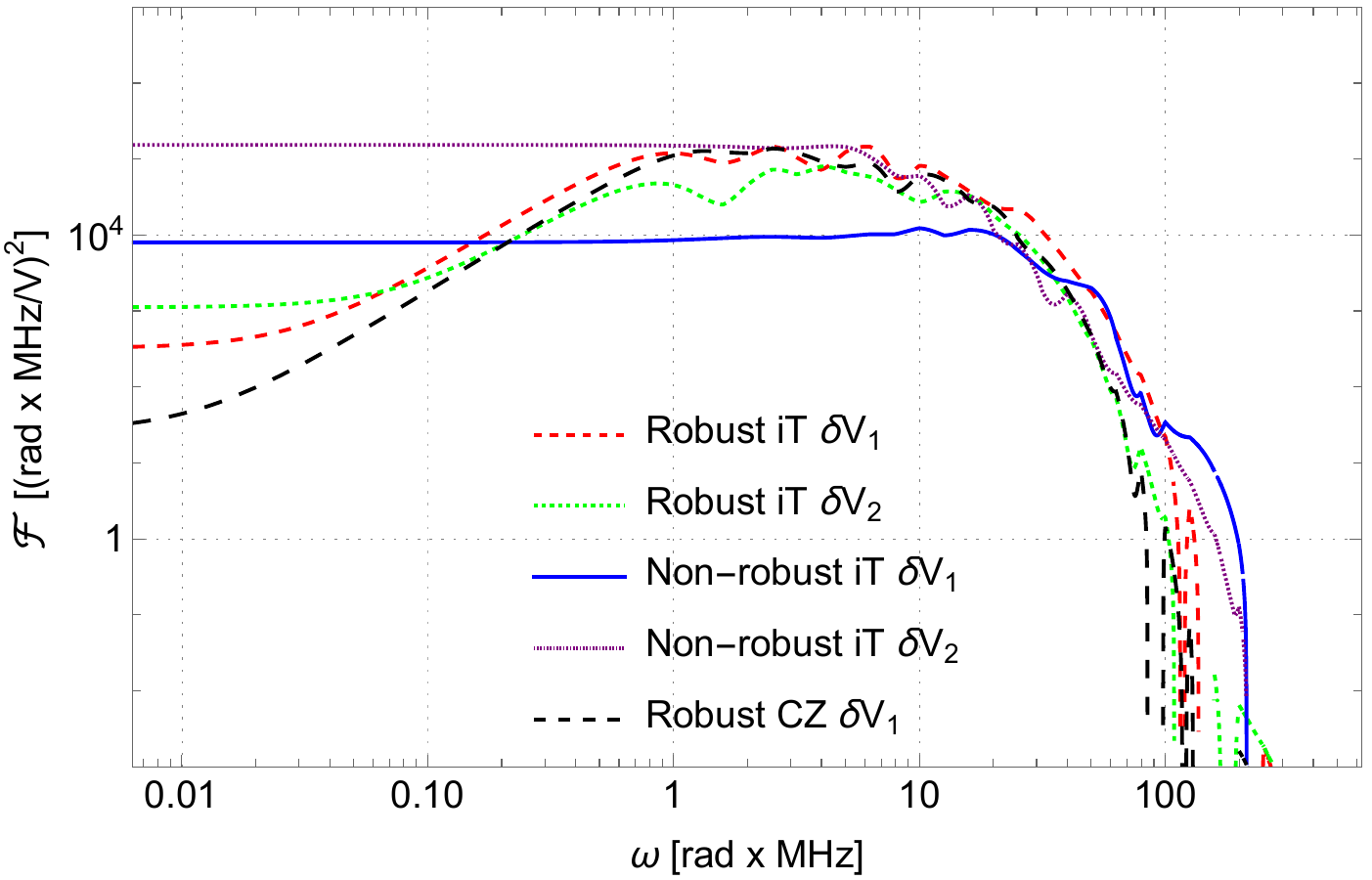}
      \caption{Top: Trace infidelity of the gates vs quasistatic fluctuations in average barrier voltage $(\delta V_1+\delta V_2)/2$. Bottom: Filter function of the neural-network designed gates for barrier gate fluctuations $\delta V_1$ and $\delta V_2$.}
      \label{fig:chargenoise}
\end{figure}

The fidelity as a function of the average quasistatic fluctuation of voltages, $(\delta V_1+\delta V_2)/2$ for all entangling gates considered are shown in Fig.~\ref{fig:chargenoise}. The {\sc cz} infidelity remains below $10^{-3}$ for average quasistatic voltage fluctuations up to $6.0$mV.
Charge noise however is not generally quasistatic but has a frequency dependence of $1/f$ between the calibration infrared cutoff, $\omega_{\text{ir}}$ and a high frequency cutoff, $\omega_{\text{cutoff}}$, after which it falls of as $1/f^2$. To analyze the effect of time dependent charge noise we use the filter function formalism from Ref.~\cite{Green_2013}. The filter function, $\mathcal{F}(\omega)$, with our definition, is the Fourier transform of the error Hamiltonian in the toggling frame which shows the frequency response of the pulse. As voltage fluctuations are treated as uncorrelated, the gates can have separate filter functions, $\mathcal{F}_i$ for $\delta V_i$, which are plotted in Fig.~\ref{fig:chargenoise}. The {\sc cz} gate is practically unaffected by fluctuations in $\delta V_2$ since $J_2$ is not driven during the pulse so only $\mathcal{F}_1$ is calculated and plotted. An estimate of the average infidelity is found by adding together the integral $\frac{1}{2\pi}\int^{\infty}_{\omega_{\text{ir}}} S(\omega)\mathcal{F}_i(\omega)\text{d}\omega$ for both error sources $\delta V_i$. The power spectral density, $S(\omega)$, of the charge noise induced voltage fluctuations is the same for both $\delta V_i$ and described by 
\begin{equation}
    S(\omega) = 
  \begin{cases}
    A_0^2/\omega  & \text{for } \omega_{\text{ir}} \leq \omega \leq \omega_{\text{cutoff}} \\
    A_0^2\omega_{\text{cutoff}}/\omega^2 & \text{for } \omega_{\text{cutoff}} \leq \omega \leq \infty
  \end{cases}
\end{equation}
where $A_0$ is the noise strength at $1$Hz. Using an estimate $A_0=1\mu V$, $\omega_{\text{cutoff}}=100$MHz\cite{connors2020chargenoise} and $\omega_{\text{ir}}\approx10^{-3}$Hz for both voltages results in an estimated infidelity as a result of frequency dependent charge noise of $8.4 \times 10^{-3}$ for the robust {\sc cz} gate.

\subsection{iToffoli gate}
\begin{figure}[t]
    \centering
    \includegraphics[width=\linewidth]{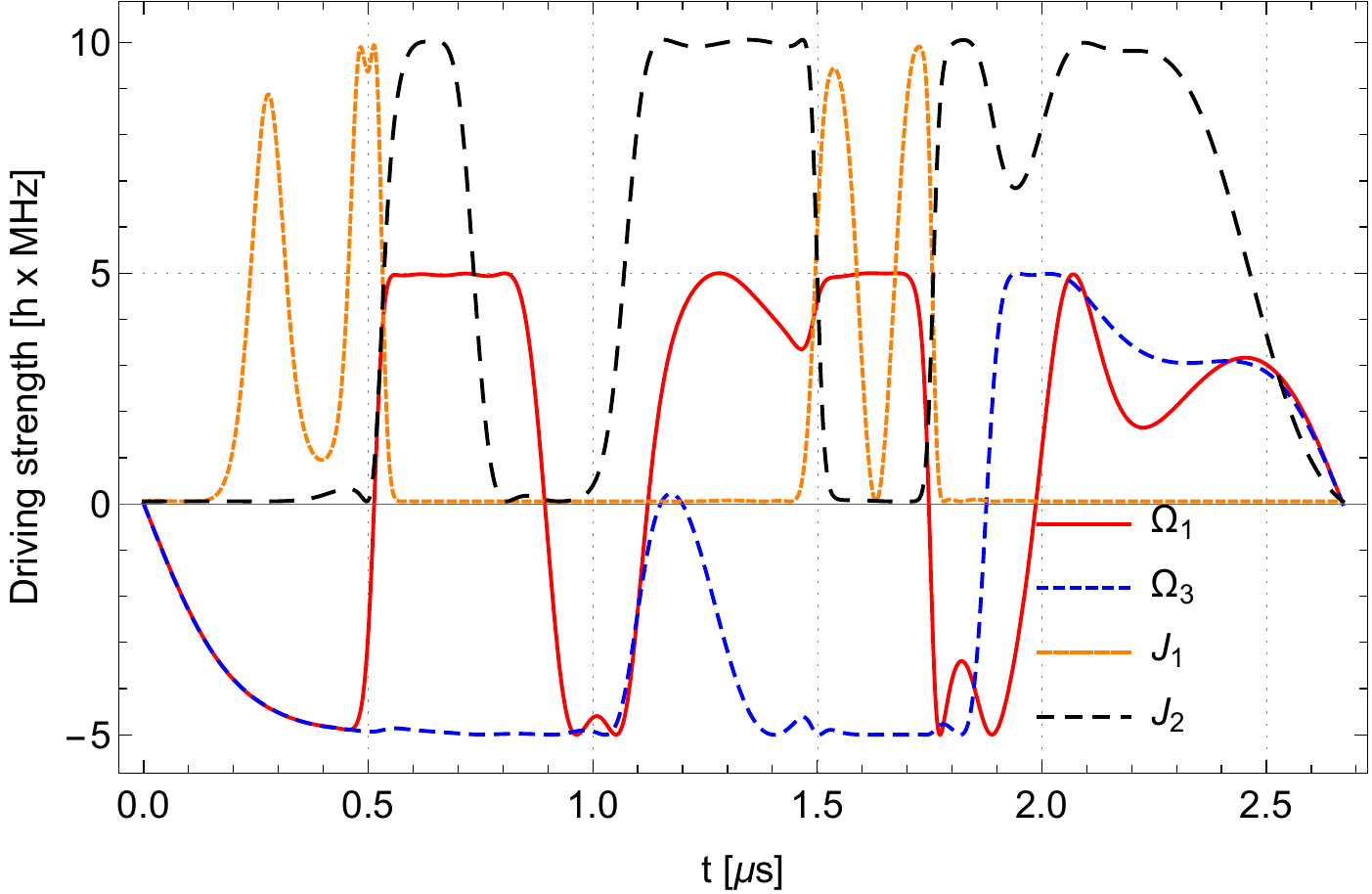}
    \includegraphics[width=\linewidth]{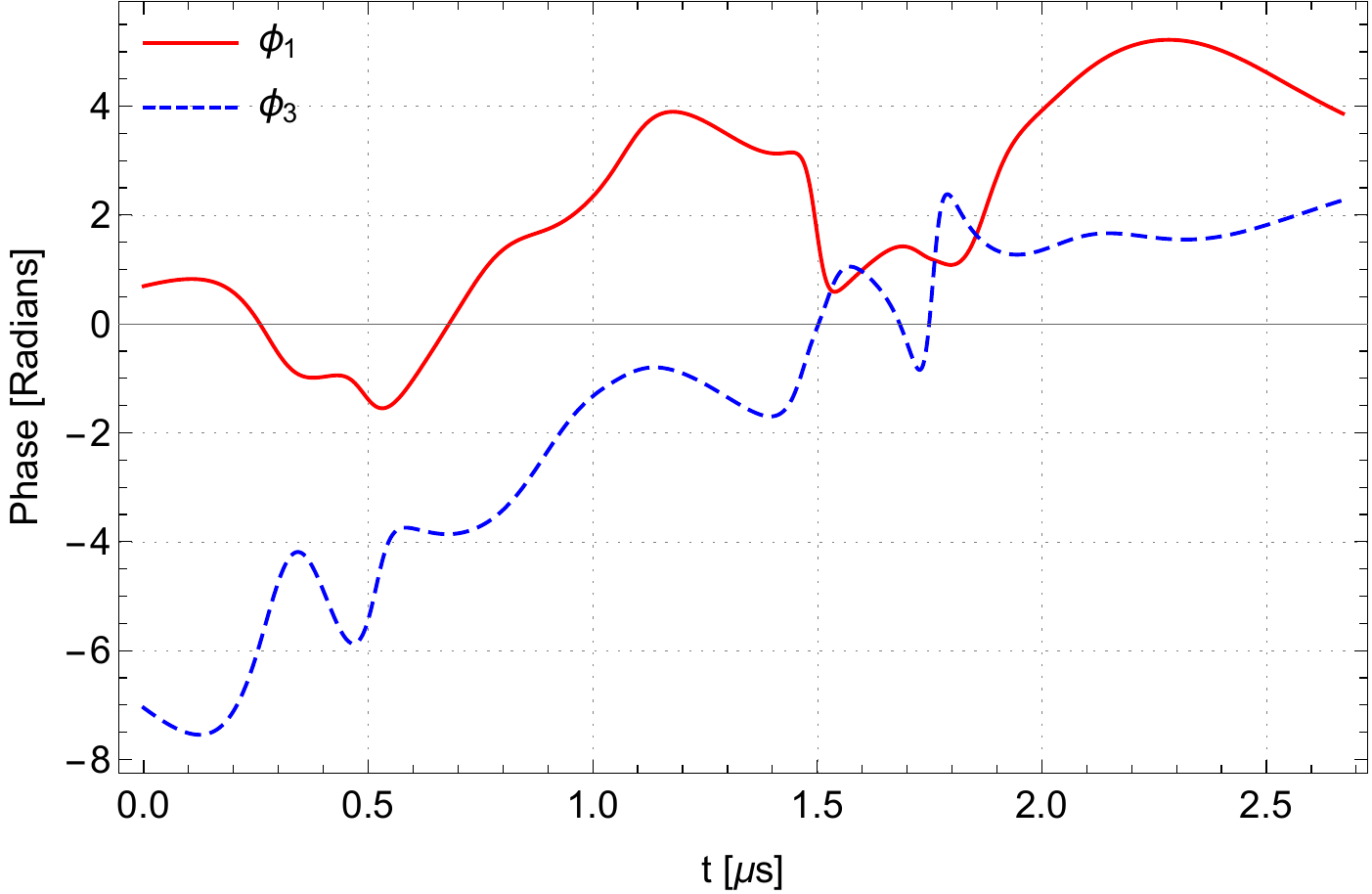}
      \caption{Neural network designed control fields vs time for a non-robust iToffoli gate in the presence of crosstalk. Top: Exchange amplitudes $J_1$ and $J_2$ as well as driving tone amplitudes $\Omega_1$ and $\Omega_3$. Bottom: Accompanying phase modulation of the driving tones $\phi_1$ and $\phi_3$.}
      \label{fig:pulseshape}
\end{figure}
\begin{figure}[t]
    \centering
    \includegraphics[width=\linewidth]{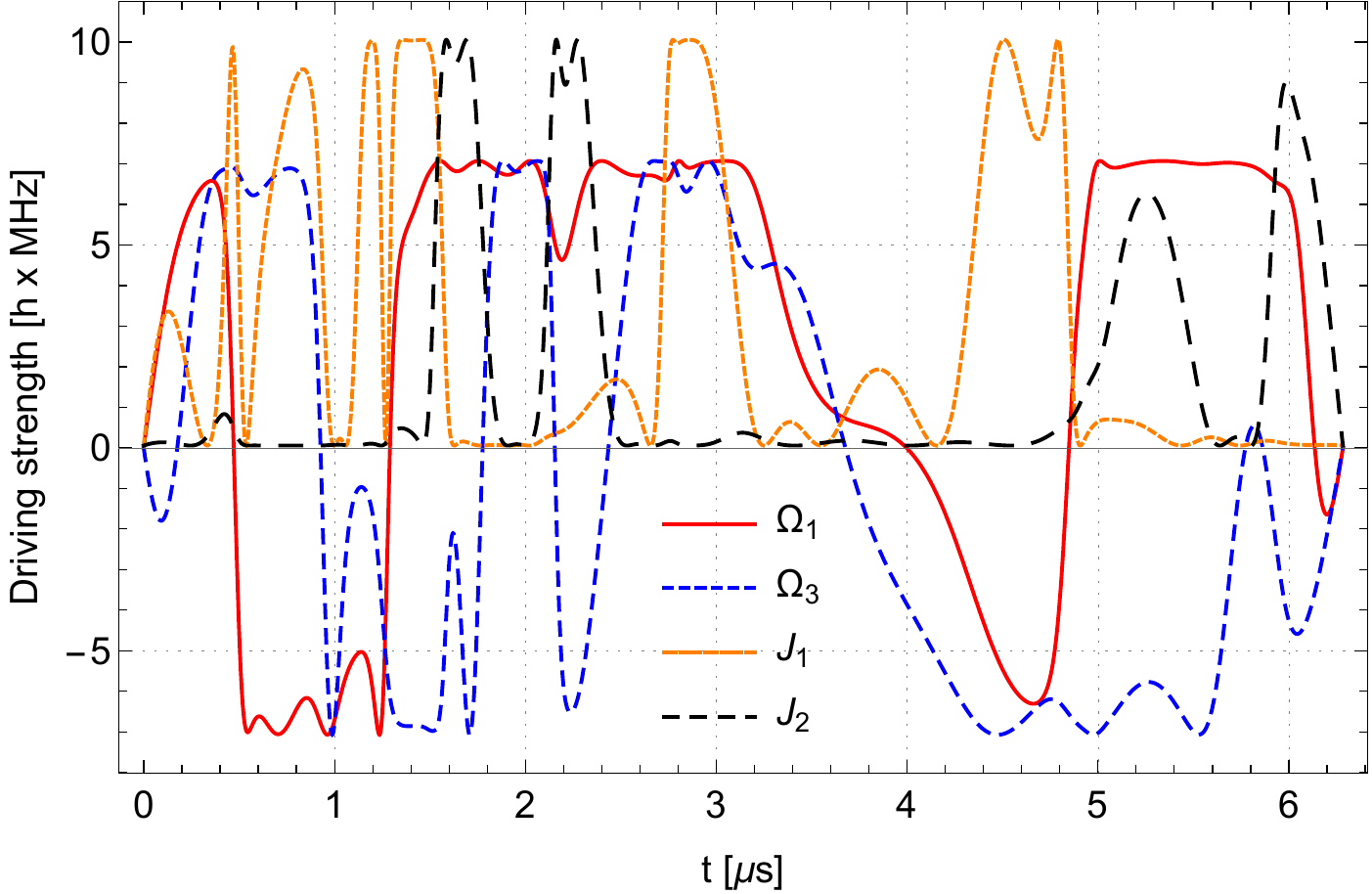}
    \includegraphics[width=\linewidth]{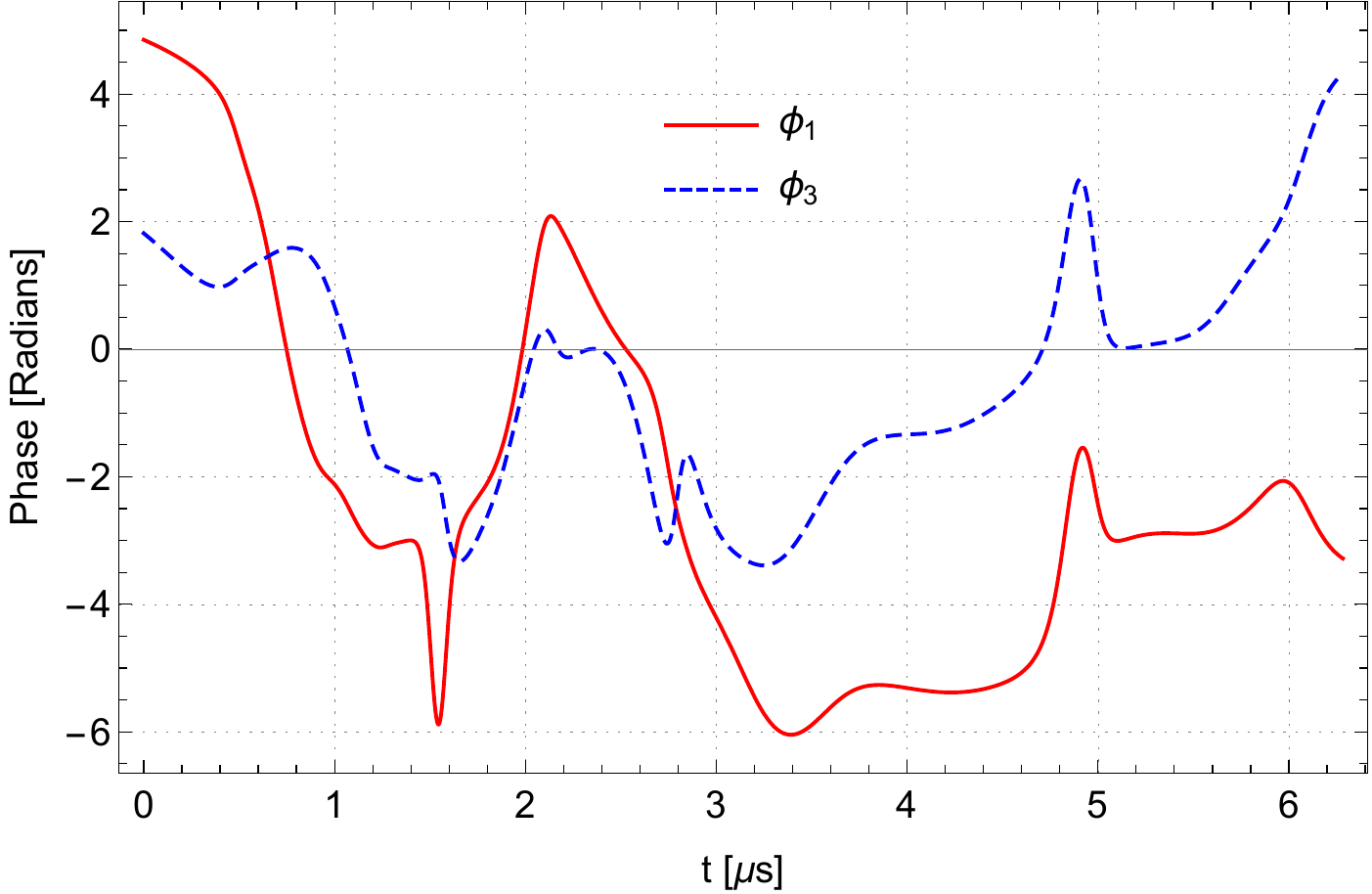}
      \caption{Neural network designed control fields vs time for charge noise robust iToffoli gate in the presence of crosstalk. Top: Exchange amplitudes $J_1$ and $J_2$ as well as driving tone amplitudes $\Omega_1$ and $\Omega_3$. Bottom: Accompanying phase modulation of the driving tones $\phi_1$ and $\phi_3$.}
      \label{fig:pulseshape2}
\end{figure}
Now we turn to the main results of this work. The control fields for the non-robust and robust iToffoli gates are shown in Figs.~\ref{fig:pulseshape} and \ref{fig:pulseshape2}. The non-robust iToffoli was implemented in a gate time of $8.5 \pi/J_{\text{max}}$ and a peak driving amplitude of $\Omega=0.71//J_{\text{max}}$, which is less than the imposed constraint $\Omega \leq \Omega_{\text{max}} = 10$MHz. The robust iToffoli used a gate time of $2 \pi/J_{\text{max}}$ and saturated the amplitude constraint.

\subsubsection{Change noise analysis}
The effect of charge noise is analyzed in the way described in section \ref{sec:chargenoiseanalysis1}. In the quasistatic case, we obtain an infidelity below $10^{-3}$ for average voltage fluctuations of 0.4mV for the non-robust pulse and 2.1mV for the robust pulse as shown in Fig.~\ref{fig:chargenoise}. To analyse the $1/f$ frequency dependent charge noise, the filter function formalism was applied again. The filter functions for both the robust and non-robust neural network designed iToffoli gates, $\mathcal{F}_i$, are plotted in Fig.~\ref{fig:chargenoise}. Using the same estimates of $A_0=1\mu V$, $\omega_{\text{cutoff}}=100$MHz\cite{connors2020chargenoise} and $\omega_{\text{ir}}\approx10^{-3}$Hz for both voltages results in an estimated infidelity as a result of frequency dependent charge noise of $1.6 \times 10^{-2}$ for the robust iToffoli, compared to $9.4 \times 10^{-2}$ for the non-robust iToffoli. The robust iToffoli infidelity is almost an order of magnitude lower and this ratio would also be maintained for devices designed to have smaller charge noise strengths \cite{paquelet_wuetz_reducing_2023}.  %For reference, if one took the same square pulses designed to produce an iToffoli in a device without crosstalk \cite{Gullans_2019} and inappropriately use them in a device that has the parameters we have used here, the average infidelity would be $0.22$ with its filter function also plotted in Fig.~\ref{fig:chargenoise}. 

\subsubsection{Bandwidth and time lengths}
When considering the pulses discussed above to mitigate charge noise, one might worry that the bandwidth of the control hardware and time length to implement the pulses can limit their usefulness. 
However, the robust iToffoli requires a 3dB bandwidth of only $1.5$MHz to implement and the non-robust iToffoli requires a 3dB bandwidth of $2.8$MHz. The robust iToffoli gate uses less bandwidth than its non-robust counterpart most likely because the non-robust iToffoli is carried out on a shorter timescale. These bandwidths are easily achievable in experiments, where the fast control lines can have bandwidths (after a low-pass filter) on the order of $200$MHz \cite{xue_cmos-based_2021}.

\begin{figure}[t]
    \centering
    \includegraphics[width=\linewidth]{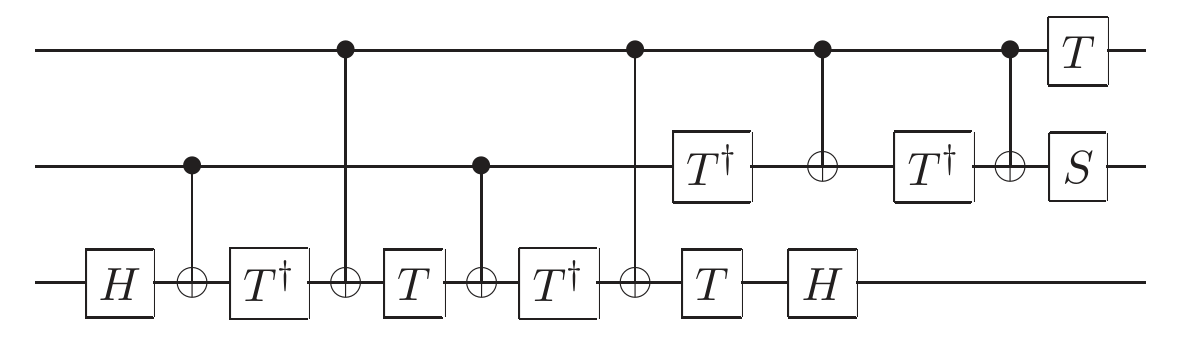}
    \includegraphics[width=0.6\linewidth]{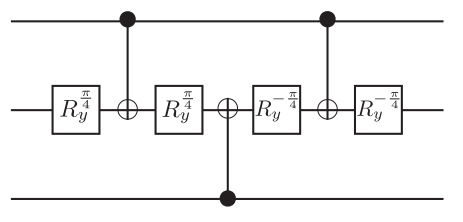}
    \caption{Top: the synthesized Toffoli gate using single and two qubit gates from Ref.~\cite{NielsenandChuang}. Here $H$ represents a Hadamard gate while $T$ and $S$ represent $\pi/8$ and $\pi/4$ phase shift gates. Bottom: The synthesized Margolus gate using single qubit $\pi/4$  $Y$ rotations, $R_{y}^{\frac{\pi}{4}}$, and two qubit CNOT gates. }
    \label{fig:Toffoli}
\end{figure}

The total time of the non-robust iToffoli pulse is $8.5 \pi /J_{\text{max}} \approx 26.7/J_{\text{max}}$ which is only slightly longer than the previously known iToffoli gate for the case where crosstalk is not an issue \cite{Gullans_2019}, which is approximately $20.6/J_{\text{max}}$ long. The length of the generated pulse is hard to compare to a Toffoli gate synthesized using one- and two-qubit gates in this three-qubit system with large crosstalk. This is because there is no straightforward way to perform the individual one- and two-qubit gates unless a similar numerical optimization is performed for each step. But to get a rough (and overly conservative) estimate, we can compare to the length of a synthesized Toffoli gate using a three-qubit system \emph{without} appreciable crosstalk. Using the same maximum driving and exchange amplitudes as in the optimization, the standard Toffoli gate synthesized as shown in Fig.~\ref{fig:Toffoli} will take $\approx 34.4/J_{\text{max}}$ if virtual-z gates are used. % and the driving has the same maximum amplitude as in our iToffoli gate. %There are six CNOT gates which each take a CZ gate and two haddamard gates as well as two individual haddamard gates. The T and S gates can be replecated using virtual-z gates and take no time. 
This is slower than the optimized pulse even in an environment where it is easier to create a gate and makes use of $ZZ$ coupling between the first and third qubit which was not directly available in our Hamiltonian. However, the gates are not completely comparable since it performs a full Toffoli gate and not a Toffoli equivalent gate. A more fair comparison would be with a Toffoli equivalent gate such as the Margolus gate acting on the middle qubit with the outer qubits as controls. The gates needed to synthesize the Margolus gate are shown in Fig.~\ref{fig:Toffoli}. Even then, the Margolus gate in a system with no crosstalk would take $20.3/J_{\text{max}}$ to complete, which is only slightly faster than the crosstalk-compensating iToffoli gate. The robust iToffoli gate took a little over twice as long as the non-robust version at $62.8/J_{\text{max}}$, but the result is better fidelity performance over a range of noise strengths.
\section{Summary \& Conclusion}
By numerically optimizing physically informed neural networks, a set of smooth control field pulses were found that result in a universal set of gates robust against charge noise in a three-qubit system with crosstalk. The gates presented in this work are a robust iToffoli, CZ, $X$ rotations and a non-robust iToffoli gate. The $X$ gates were corrected for crosstalk and were automatically robust against fluctuations in the exchange $J_i$ caused by charge noise because they required no entangling. The robust CZ, robust iToffoli and non-robust iToffoli required more complicated control but maintained an infidelity of $10^{-3}$ up to quasistatic average fluctuation of voltage of 6.0mV, 2.1mV, and 0.4mV (respectively) which cause exchange fluctuations through the exponential dependence of $J$ on $V$. For $1/f$ frequency-dependent charge noise, similar order-of-magnitude reductions in infidelity were obtained from the robust shaped pulse iToffoli performed better than non-robust shaped and square pulse iToffolis. The robust pulse required only modest bandwidth and a time comparable to the most efficient non-robust iToffoli gates, including those that only apply to systems without crosstalk.
\section*{Acknowledgements} \label{sec:acknowledgements}
    This research was sponsored by the Army Research Office (ARO), and was accomplished under Grant Number W911NF-17-1-0287.

\bibliographystyle{apsrev4-1} 
\bibliography{refs} 

\begin{thebibliography}{35}%
\makeatletter
\providecommand \@ifxundefined [1]{%
 \@ifx{#1\undefined}
}%
\providecommand \@ifnum [1]{%
 \ifnum #1\expandafter \@firstoftwo
 \else \expandafter \@secondoftwo
 \fi
}%
\providecommand \@ifx [1]{%
 \ifx #1\expandafter \@firstoftwo
 \else \expandafter \@secondoftwo
 \fi
}%
\providecommand \natexlab [1]{#1}%
\providecommand \enquote  [1]{``#1''}%
\providecommand \bibnamefont  [1]{#1}%
\providecommand \bibfnamefont [1]{#1}%
\providecommand \citenamefont [1]{#1}%
\providecommand \href@noop [0]{\@secondoftwo}%
\providecommand \href [0]{\begingroup \@sanitize@url \@href}%
\providecommand \@href[1]{\@@startlink{#1}\@@href}%
\providecommand \@@href[1]{\endgroup#1\@@endlink}%
\providecommand \@sanitize@url [0]{\catcode `\\12\catcode `\$12\catcode
  `\&12\catcode `\#12\catcode `\^12\catcode `\_12\catcode `\%12\relax}%
\providecommand \@@startlink[1]{}%
\providecommand \@@endlink[0]{}%
\providecommand \url  [0]{\begingroup\@sanitize@url \@url }%
\providecommand \@url [1]{\endgroup\@href {#1}{\urlprefix }}%
\providecommand \urlprefix  [0]{URL }%
\providecommand \Eprint [0]{\href }%
\providecommand \doibase [0]{http://dx.doi.org/}%
\providecommand \selectlanguage [0]{\@gobble}%
\providecommand \bibinfo  [0]{\@secondoftwo}%
\providecommand \bibfield  [0]{\@secondoftwo}%
\providecommand \translation [1]{[#1]}%
\providecommand \BibitemOpen [0]{}%
\providecommand \bibitemStop [0]{}%
\providecommand \bibitemNoStop [0]{.\EOS\space}%
\providecommand \EOS [0]{\spacefactor3000\relax}%
\providecommand \BibitemShut  [1]{\csname bibitem#1\endcsname}%
\let\auto@bib@innerbib\@empty
%</preamble>
\bibitem [{\citenamefont {Yang}\ \emph {et~al.}(2019)\citenamefont {Yang},
  \citenamefont {Chan}, \citenamefont {Harper}, \citenamefont {Huang},
  \citenamefont {Evans}, \citenamefont {Hwang}, \citenamefont {Hensen},
  \citenamefont {Laucht}, \citenamefont {Tanttu}, \citenamefont {Hudson},
  \citenamefont {Flammia}, \citenamefont {Itoh}, \citenamefont {Morello},
  \citenamefont {Bartlett},\ and\ \citenamefont {Dzurak}}]{Yang2019}%
  \BibitemOpen
  \bibfield  {author} {\bibinfo {author} {\bibfnamefont {C.~H.}\ \bibnamefont
  {Yang}}, \bibinfo {author} {\bibfnamefont {K.~W.}\ \bibnamefont {Chan}},
  \bibinfo {author} {\bibfnamefont {R.}~\bibnamefont {Harper}}, \bibinfo
  {author} {\bibfnamefont {W.}~\bibnamefont {Huang}}, \bibinfo {author}
  {\bibfnamefont {T.}~\bibnamefont {Evans}}, \bibinfo {author} {\bibfnamefont
  {J.~C.}\ \bibnamefont {Hwang}}, \bibinfo {author} {\bibfnamefont
  {B.}~\bibnamefont {Hensen}}, \bibinfo {author} {\bibfnamefont
  {A.}~\bibnamefont {Laucht}}, \bibinfo {author} {\bibfnamefont
  {T.}~\bibnamefont {Tanttu}}, \bibinfo {author} {\bibfnamefont {F.~E.}\
  \bibnamefont {Hudson}}, \bibinfo {author} {\bibfnamefont {S.~T.}\
  \bibnamefont {Flammia}}, \bibinfo {author} {\bibfnamefont {K.~M.}\
  \bibnamefont {Itoh}}, \bibinfo {author} {\bibfnamefont {A.}~\bibnamefont
  {Morello}}, \bibinfo {author} {\bibfnamefont {S.~D.}\ \bibnamefont
  {Bartlett}}, \ and\ \bibinfo {author} {\bibfnamefont {A.~S.}\ \bibnamefont
  {Dzurak}},\ }\href {\doibase 10.1038/s41928-019-0234-1} {\bibfield  {journal}
  {\bibinfo  {journal} {Nat. Electron.}\ }\textbf {\bibinfo {volume} {2}},\
  \bibinfo {pages} {151} (\bibinfo {year} {2019})}\BibitemShut {NoStop}%
\bibitem [{\citenamefont {Yoneda}\ \emph {et~al.}(2018)\citenamefont {Yoneda},
  \citenamefont {Takeda}, \citenamefont {Otsuka}, \citenamefont {Nakajima},
  \citenamefont {Delbecq}, \citenamefont {Allison}, \citenamefont {Honda},
  \citenamefont {Kodera}, \citenamefont {Oda}, \citenamefont {Hoshi},
  \citenamefont {Usami}, \citenamefont {Itoh},\ and\ \citenamefont
  {Tarucha}}]{Yoneda2018}%
  \BibitemOpen
  \bibfield  {author} {\bibinfo {author} {\bibfnamefont {J.}~\bibnamefont
  {Yoneda}}, \bibinfo {author} {\bibfnamefont {K.}~\bibnamefont {Takeda}},
  \bibinfo {author} {\bibfnamefont {T.}~\bibnamefont {Otsuka}}, \bibinfo
  {author} {\bibfnamefont {T.}~\bibnamefont {Nakajima}}, \bibinfo {author}
  {\bibfnamefont {M.~R.}\ \bibnamefont {Delbecq}}, \bibinfo {author}
  {\bibfnamefont {G.}~\bibnamefont {Allison}}, \bibinfo {author} {\bibfnamefont
  {T.}~\bibnamefont {Honda}}, \bibinfo {author} {\bibfnamefont
  {T.}~\bibnamefont {Kodera}}, \bibinfo {author} {\bibfnamefont
  {S.}~\bibnamefont {Oda}}, \bibinfo {author} {\bibfnamefont {Y.}~\bibnamefont
  {Hoshi}}, \bibinfo {author} {\bibfnamefont {N.}~\bibnamefont {Usami}},
  \bibinfo {author} {\bibfnamefont {K.~M.}\ \bibnamefont {Itoh}}, \ and\
  \bibinfo {author} {\bibfnamefont {S.}~\bibnamefont {Tarucha}},\ }\href
  {http://dx.doi.org/10.1038/s41565-017-0014-x} {\bibfield  {journal} {\bibinfo
   {journal} {Nat. Nanotechnol.}\ }\textbf {\bibinfo {volume} {13}},\ \bibinfo
  {pages} {102–106} (\bibinfo {year} {2018})}\BibitemShut {NoStop}%
\bibitem [{\citenamefont {Petit}\ \emph {et~al.}(2020)\citenamefont {Petit},
  \citenamefont {Eenink}, \citenamefont {Russ}, \citenamefont {Lawrie},
  \citenamefont {Hendrickx}, \citenamefont {Philips}, \citenamefont {Clarke},
  \citenamefont {Vandersypen},\ and\ \citenamefont {Veldhorst}}]{Petit_2020}%
  \BibitemOpen
  \bibfield  {author} {\bibinfo {author} {\bibfnamefont {L.}~\bibnamefont
  {Petit}}, \bibinfo {author} {\bibfnamefont {H.~G.}\ \bibnamefont {Eenink}},
  \bibinfo {author} {\bibfnamefont {M.}~\bibnamefont {Russ}}, \bibinfo {author}
  {\bibfnamefont {W.~I.}\ \bibnamefont {Lawrie}}, \bibinfo {author}
  {\bibfnamefont {N.~W.}\ \bibnamefont {Hendrickx}}, \bibinfo {author}
  {\bibfnamefont {S.~G.}\ \bibnamefont {Philips}}, \bibinfo {author}
  {\bibfnamefont {J.~S.}\ \bibnamefont {Clarke}}, \bibinfo {author}
  {\bibfnamefont {L.~M.}\ \bibnamefont {Vandersypen}}, \ and\ \bibinfo {author}
  {\bibfnamefont {M.}~\bibnamefont {Veldhorst}},\ }\href {\doibase
  10.1038/s41586-020-2170-7} {\bibfield  {journal} {\bibinfo  {journal}
  {Nature}\ }\textbf {\bibinfo {volume} {580}},\ \bibinfo {pages} {355}
  (\bibinfo {year} {2020})}\BibitemShut {NoStop}%
\bibitem [{\citenamefont {Xue}\ \emph {et~al.}(2022)\citenamefont {Xue},
  \citenamefont {Russ}, \citenamefont {Samkharadze}, \citenamefont {Undseth},
  \citenamefont {Sammak}, \citenamefont {Scappucci},\ and\ \citenamefont
  {Vandersypen}}]{xue2021}%
  \BibitemOpen
  \bibfield  {author} {\bibinfo {author} {\bibfnamefont {X.}~\bibnamefont
  {Xue}}, \bibinfo {author} {\bibfnamefont {M.}~\bibnamefont {Russ}}, \bibinfo
  {author} {\bibfnamefont {N.}~\bibnamefont {Samkharadze}}, \bibinfo {author}
  {\bibfnamefont {B.}~\bibnamefont {Undseth}}, \bibinfo {author} {\bibfnamefont
  {A.}~\bibnamefont {Sammak}}, \bibinfo {author} {\bibfnamefont
  {G.}~\bibnamefont {Scappucci}}, \ and\ \bibinfo {author} {\bibfnamefont
  {L.~M.}\ \bibnamefont {Vandersypen}},\ }\href {\doibase
  10.1038/s41586-021-04273-w} {\bibfield  {journal} {\bibinfo  {journal}
  {Nature}\ }\textbf {\bibinfo {volume} {601}},\ \bibinfo {pages} {343–347}
  (\bibinfo {year} {2022})}\BibitemShut {NoStop}%
\bibitem [{\citenamefont {Noiri}\ \emph {et~al.}(2022)\citenamefont {Noiri},
  \citenamefont {Takeda}, \citenamefont {Nakajima}, \citenamefont {Kobayashi},
  \citenamefont {Sammak}, \citenamefont {Scappucci},\ and\ \citenamefont
  {Tarucha}}]{Noiri2022}%
  \BibitemOpen
  \bibfield  {author} {\bibinfo {author} {\bibfnamefont {A.}~\bibnamefont
  {Noiri}}, \bibinfo {author} {\bibfnamefont {K.}~\bibnamefont {Takeda}},
  \bibinfo {author} {\bibfnamefont {T.}~\bibnamefont {Nakajima}}, \bibinfo
  {author} {\bibfnamefont {T.}~\bibnamefont {Kobayashi}}, \bibinfo {author}
  {\bibfnamefont {A.}~\bibnamefont {Sammak}}, \bibinfo {author} {\bibfnamefont
  {G.}~\bibnamefont {Scappucci}}, \ and\ \bibinfo {author} {\bibfnamefont
  {S.}~\bibnamefont {Tarucha}},\ }\href {\doibase 10.1038/s41586-021-04182-y}
  {\bibfield  {journal} {\bibinfo  {journal} {Nature}\ }\textbf {\bibinfo
  {volume} {601}},\ \bibinfo {pages} {338} (\bibinfo {year}
  {2022})}\BibitemShut {NoStop}%
\bibitem [{\citenamefont {Mills}\ \emph {et~al.}(2022)\citenamefont {Mills},
  \citenamefont {Guinn}, \citenamefont {Gullans}, \citenamefont {Sigillito},
  \citenamefont {Feldman}, \citenamefont {Nielsen},\ and\ \citenamefont
  {Petta}}]{mills2021twoqubit}%
  \BibitemOpen
  \bibfield  {author} {\bibinfo {author} {\bibfnamefont {A.~R.}\ \bibnamefont
  {Mills}}, \bibinfo {author} {\bibfnamefont {C.~R.}\ \bibnamefont {Guinn}},
  \bibinfo {author} {\bibfnamefont {M.~J.}\ \bibnamefont {Gullans}}, \bibinfo
  {author} {\bibfnamefont {A.~J.}\ \bibnamefont {Sigillito}}, \bibinfo {author}
  {\bibfnamefont {M.~M.}\ \bibnamefont {Feldman}}, \bibinfo {author}
  {\bibfnamefont {E.}~\bibnamefont {Nielsen}}, \ and\ \bibinfo {author}
  {\bibfnamefont {J.~R.}\ \bibnamefont {Petta}},\ }\href {\doibase
  10.1126/sciadv.abn5130} {\bibfield  {journal} {\bibinfo  {journal} {Science
  Advances}\ }\textbf {\bibinfo {volume} {8}},\ \bibinfo {pages} {eabn5130}
  (\bibinfo {year} {2022})}\BibitemShut {NoStop}%
\bibitem [{\citenamefont {Huang}\ \emph {et~al.}(2019)\citenamefont {Huang},
  \citenamefont {Yang}, \citenamefont {Chan}, \citenamefont {Tanttu},
  \citenamefont {Hensen}, \citenamefont {Leon}, \citenamefont {Fogarty},
  \citenamefont {Hwang}, \citenamefont {Hudson}, \citenamefont {Itoh},
  \citenamefont {Morello}, \citenamefont {Laucht},\ and\ \citenamefont
  {Dzurak}}]{Huang2019}%
  \BibitemOpen
  \bibfield  {author} {\bibinfo {author} {\bibfnamefont {W.}~\bibnamefont
  {Huang}}, \bibinfo {author} {\bibfnamefont {C.~H.}\ \bibnamefont {Yang}},
  \bibinfo {author} {\bibfnamefont {K.~W.}\ \bibnamefont {Chan}}, \bibinfo
  {author} {\bibfnamefont {T.}~\bibnamefont {Tanttu}}, \bibinfo {author}
  {\bibfnamefont {B.}~\bibnamefont {Hensen}}, \bibinfo {author} {\bibfnamefont
  {R.~C.}\ \bibnamefont {Leon}}, \bibinfo {author} {\bibfnamefont {M.~A.}\
  \bibnamefont {Fogarty}}, \bibinfo {author} {\bibfnamefont {J.~C.}\
  \bibnamefont {Hwang}}, \bibinfo {author} {\bibfnamefont {F.~E.}\ \bibnamefont
  {Hudson}}, \bibinfo {author} {\bibfnamefont {K.~M.}\ \bibnamefont {Itoh}},
  \bibinfo {author} {\bibfnamefont {A.}~\bibnamefont {Morello}}, \bibinfo
  {author} {\bibfnamefont {A.}~\bibnamefont {Laucht}}, \ and\ \bibinfo {author}
  {\bibfnamefont {A.~S.}\ \bibnamefont {Dzurak}},\ }\href {\doibase
  10.1038/s41586-019-1197-0} {\bibfield  {journal} {\bibinfo  {journal}
  {Nature}\ }\textbf {\bibinfo {volume} {569}},\ \bibinfo {pages} {532}
  (\bibinfo {year} {2019})}\BibitemShut {NoStop}%
\bibitem [{\citenamefont {Fowler}\ \emph {et~al.}(2012)\citenamefont {Fowler},
  \citenamefont {Mariantoni}, \citenamefont {Martinis},\ and\ \citenamefont
  {Cleland}}]{Fowler2012}%
  \BibitemOpen
  \bibfield  {author} {\bibinfo {author} {\bibfnamefont {A.~G.}\ \bibnamefont
  {Fowler}}, \bibinfo {author} {\bibfnamefont {M.}~\bibnamefont {Mariantoni}},
  \bibinfo {author} {\bibfnamefont {J.~M.}\ \bibnamefont {Martinis}}, \ and\
  \bibinfo {author} {\bibfnamefont {A.~N.}\ \bibnamefont {Cleland}},\ }\href
  {\doibase 10.1103/PhysRevA.86.032324} {\bibfield  {journal} {\bibinfo
  {journal} {Phys. Rev. A}\ }\textbf {\bibinfo {volume} {86}},\ \bibinfo
  {pages} {032324} (\bibinfo {year} {2012})}\BibitemShut {NoStop}%
\bibitem [{\citenamefont {Nielsen}\ and\ \citenamefont
  {Chuang}(2011)}]{NielsenandChuang}%
  \BibitemOpen
  \bibfield  {author} {\bibinfo {author} {\bibfnamefont {M.~A.}\ \bibnamefont
  {Nielsen}}\ and\ \bibinfo {author} {\bibfnamefont {I.~L.}\ \bibnamefont
  {Chuang}},\ }\href@noop {} {\emph {\bibinfo {title} {Quantum Computation and
  Quantum Information: 10th Anniversary Edition}}},\ \bibinfo {edition} {10th}\
  ed.\ (\bibinfo  {publisher} {Cambridge University Press},\ \bibinfo {address}
  {USA},\ \bibinfo {year} {2011})\BibitemShut {NoStop}%
\bibitem [{\citenamefont {Kitaev}(1997)}]{Kitaev1997}%
  \BibitemOpen
  \bibfield  {author} {\bibinfo {author} {\bibfnamefont {A.~Y.}\ \bibnamefont
  {Kitaev}},\ }\href {\doibase 10.1070/RM1997v052n06ABEH002155} {\bibfield
  {journal} {\bibinfo  {journal} {Russ. Math. Surv.}\ }\textbf {\bibinfo
  {volume} {52}},\ \bibinfo {pages} {1191} (\bibinfo {year}
  {1997})}\BibitemShut {NoStop}%
\bibitem [{\citenamefont {Shende}\ and\ \citenamefont
  {Markov}(2009)}]{Shende2008}%
  \BibitemOpen
  \bibfield  {author} {\bibinfo {author} {\bibfnamefont {V.~V.}\ \bibnamefont
  {Shende}}\ and\ \bibinfo {author} {\bibfnamefont {I.~L.}\ \bibnamefont
  {Markov}},\ }\href {https://dl.acm.org/doi/10.5555/2011791.2011799}
  {\bibfield  {journal} {\bibinfo  {journal} {Quant. Inf. Comp.}\ }\textbf
  {\bibinfo {volume} {9}},\ \bibinfo {pages} {461–486} (\bibinfo {year}
  {2009})}\BibitemShut {NoStop}%
\bibitem [{\citenamefont {Maslov}(2016)}]{Maslov2016}%
  \BibitemOpen
  \bibfield  {author} {\bibinfo {author} {\bibfnamefont {D.}~\bibnamefont
  {Maslov}},\ }\href {\doibase 10.1103/PhysRevA.93.022311} {\bibfield
  {journal} {\bibinfo  {journal} {Phys. Rev. A}\ }\textbf {\bibinfo {volume}
  {93}},\ \bibinfo {pages} {022311} (\bibinfo {year} {2016})}\BibitemShut
  {NoStop}%
\bibitem [{\citenamefont {Gullans}\ and\ \citenamefont
  {Petta}(2019)}]{Gullans_2019}%
  \BibitemOpen
  \bibfield  {author} {\bibinfo {author} {\bibfnamefont {M.~J.}\ \bibnamefont
  {Gullans}}\ and\ \bibinfo {author} {\bibfnamefont {J.~R.}\ \bibnamefont
  {Petta}},\ }\href {https://link.aps.org/doi/10.1103/PhysRevB.100.085419}
  {\bibfield  {journal} {\bibinfo  {journal} {Phys. Rev. B}\ }\textbf {\bibinfo
  {volume} {100}},\ \bibinfo {pages} {085419} (\bibinfo {year}
  {2019})}\BibitemShut {NoStop}%
\bibitem [{\citenamefont {Seedhouse}\ \emph {et~al.}(2021)\citenamefont
  {Seedhouse}, \citenamefont {Hansen}, \citenamefont {Laucht}, \citenamefont
  {Yang}, \citenamefont {Dzurak},\ and\ \citenamefont
  {Saraiva}}]{seedhouse2021}%
  \BibitemOpen
  \bibfield  {author} {\bibinfo {author} {\bibfnamefont {A.~E.}\ \bibnamefont
  {Seedhouse}}, \bibinfo {author} {\bibfnamefont {I.}~\bibnamefont {Hansen}},
  \bibinfo {author} {\bibfnamefont {A.}~\bibnamefont {Laucht}}, \bibinfo
  {author} {\bibfnamefont {C.~H.}\ \bibnamefont {Yang}}, \bibinfo {author}
  {\bibfnamefont {A.~S.}\ \bibnamefont {Dzurak}}, \ and\ \bibinfo {author}
  {\bibfnamefont {A.}~\bibnamefont {Saraiva}},\ }\href {\doibase
  10.1103/PhysRevB.104.235411} {\bibfield  {journal} {\bibinfo  {journal}
  {Phys. Rev. B}\ }\textbf {\bibinfo {volume} {104}},\ \bibinfo {pages}
  {235411} (\bibinfo {year} {2021})}\BibitemShut {NoStop}%
\bibitem [{\citenamefont {Hansen}\ \emph {et~al.}(2021)\citenamefont {Hansen},
  \citenamefont {Seedhouse}, \citenamefont {Saraiva}, \citenamefont {Laucht},
  \citenamefont {Dzurak},\ and\ \citenamefont {Yang}}]{Hansen2021}%
  \BibitemOpen
  \bibfield  {author} {\bibinfo {author} {\bibfnamefont {I.}~\bibnamefont
  {Hansen}}, \bibinfo {author} {\bibfnamefont {A.~E.}\ \bibnamefont
  {Seedhouse}}, \bibinfo {author} {\bibfnamefont {A.}~\bibnamefont {Saraiva}},
  \bibinfo {author} {\bibfnamefont {A.}~\bibnamefont {Laucht}}, \bibinfo
  {author} {\bibfnamefont {A.~S.}\ \bibnamefont {Dzurak}}, \ and\ \bibinfo
  {author} {\bibfnamefont {C.~H.}\ \bibnamefont {Yang}},\ }\href {\doibase
  10.1103/PhysRevA.104.062415} {\bibfield  {journal} {\bibinfo  {journal}
  {Phys. Rev. A}\ }\textbf {\bibinfo {volume} {104}},\ \bibinfo {pages}
  {062415} (\bibinfo {year} {2021})}\BibitemShut {NoStop}%
\bibitem [{\citenamefont {Heinz}\ and\ \citenamefont
  {Burkard}(2021)}]{Heinz2021}%
  \BibitemOpen
  \bibfield  {author} {\bibinfo {author} {\bibfnamefont {I.}~\bibnamefont
  {Heinz}}\ and\ \bibinfo {author} {\bibfnamefont {G.}~\bibnamefont
  {Burkard}},\ }\href {\doibase 10.1103/PhysRevB.104.045420} {\bibfield
  {journal} {\bibinfo  {journal} {Phys. Rev. B}\ }\textbf {\bibinfo {volume}
  {104}},\ \bibinfo {pages} {045420} (\bibinfo {year} {2021})}\BibitemShut
  {NoStop}%
\bibitem [{\citenamefont {Huang}\ \emph {et~al.}(2018)\citenamefont {Huang},
  \citenamefont {Zimmerman},\ and\ \citenamefont {Bryant}}]{huang_spin_2018}%
  \BibitemOpen
  \bibfield  {author} {\bibinfo {author} {\bibfnamefont {P.}~\bibnamefont
  {Huang}}, \bibinfo {author} {\bibfnamefont {N.~M.}\ \bibnamefont
  {Zimmerman}}, \ and\ \bibinfo {author} {\bibfnamefont {G.~W.}\ \bibnamefont
  {Bryant}},\ }\href {\doibase 10.1038/s41534-018-0112-0} {\bibfield  {journal}
  {\bibinfo  {journal} {npj Quantum Inf}\ }\textbf {\bibinfo {volume} {4}},\
  \bibinfo {pages} {62} (\bibinfo {year} {2018})}\BibitemShut {NoStop}%
\bibitem [{\citenamefont {van Dijk}\ \emph {et~al.}(2019)\citenamefont {van
  Dijk}, \citenamefont {Kawakami}, \citenamefont {Schouten}, \citenamefont
  {Veldhorst}, \citenamefont {Vandersypen}, \citenamefont {Babaie},
  \citenamefont {Charbon},\ and\ \citenamefont {Sebastiano}}]{vanDijk2019}%
  \BibitemOpen
  \bibfield  {author} {\bibinfo {author} {\bibfnamefont {J.}~\bibnamefont {van
  Dijk}}, \bibinfo {author} {\bibfnamefont {E.}~\bibnamefont {Kawakami}},
  \bibinfo {author} {\bibfnamefont {R.}~\bibnamefont {Schouten}}, \bibinfo
  {author} {\bibfnamefont {M.}~\bibnamefont {Veldhorst}}, \bibinfo {author}
  {\bibfnamefont {L.}~\bibnamefont {Vandersypen}}, \bibinfo {author}
  {\bibfnamefont {M.}~\bibnamefont {Babaie}}, \bibinfo {author} {\bibfnamefont
  {E.}~\bibnamefont {Charbon}}, \ and\ \bibinfo {author} {\bibfnamefont
  {F.}~\bibnamefont {Sebastiano}},\ }\href {\doibase
  10.1103/PhysRevApplied.12.044054} {\bibfield  {journal} {\bibinfo  {journal}
  {Phys. Rev. Applied}\ }\textbf {\bibinfo {volume} {12}},\ \bibinfo {pages}
  {044054} (\bibinfo {year} {2019})}\BibitemShut {NoStop}%
\bibitem [{\citenamefont {G\"ung\"ord\"u}\ and\ \citenamefont
  {Kestner}(2020)}]{Gungordu2020}%
  \BibitemOpen
  \bibfield  {author} {\bibinfo {author} {\bibfnamefont {U.}~\bibnamefont
  {G\"ung\"ord\"u}}\ and\ \bibinfo {author} {\bibfnamefont {J.~P.}\
  \bibnamefont {Kestner}},\ }\href {\doibase 10.1103/PhysRevB.101.155301}
  {\bibfield  {journal} {\bibinfo  {journal} {Phys. Rev. B}\ }\textbf {\bibinfo
  {volume} {101}},\ \bibinfo {pages} {155301} (\bibinfo {year}
  {2020})}\BibitemShut {NoStop}%
\bibitem [{\citenamefont {G\"ung\"ord\"u}\ and\ \citenamefont
  {Kestner}(2022)}]{Gungordu2020p2}%
  \BibitemOpen
  \bibfield  {author} {\bibinfo {author} {\bibfnamefont {U.}~\bibnamefont
  {G\"ung\"ord\"u}}\ and\ \bibinfo {author} {\bibfnamefont {J.~P.}\
  \bibnamefont {Kestner}},\ }\href {\doibase 10.1103/PhysRevResearch.4.023155}
  {\bibfield  {journal} {\bibinfo  {journal} {Phys. Rev. Research}\ }\textbf
  {\bibinfo {volume} {4}},\ \bibinfo {pages} {023155} (\bibinfo {year}
  {2022})}\BibitemShut {NoStop}%
\bibitem [{\citenamefont {Kanaar}\ \emph {et~al.}(2022)\citenamefont {Kanaar},
  \citenamefont {G\"ung\"ord\"u},\ and\ \citenamefont {Kestner}}]{Kanaar2022}%
  \BibitemOpen
  \bibfield  {author} {\bibinfo {author} {\bibfnamefont {D.~W.}\ \bibnamefont
  {Kanaar}}, \bibinfo {author} {\bibfnamefont {U.}~\bibnamefont
  {G\"ung\"ord\"u}}, \ and\ \bibinfo {author} {\bibfnamefont {J.~P.}\
  \bibnamefont {Kestner}},\ }\href {\doibase 10.1103/PhysRevB.105.245308}
  {\bibfield  {journal} {\bibinfo  {journal} {Phys. Rev. B}\ }\textbf {\bibinfo
  {volume} {105}},\ \bibinfo {pages} {245308} (\bibinfo {year}
  {2022})}\BibitemShut {NoStop}%
\bibitem [{\citenamefont {McKay}\ \emph {et~al.}(2017)\citenamefont {McKay},
  \citenamefont {Wood}, \citenamefont {Sheldon}, \citenamefont {Chow},\ and\
  \citenamefont {Gambetta}}]{Mckay2017}%
  \BibitemOpen
  \bibfield  {author} {\bibinfo {author} {\bibfnamefont {D.~C.}\ \bibnamefont
  {McKay}}, \bibinfo {author} {\bibfnamefont {C.~J.}\ \bibnamefont {Wood}},
  \bibinfo {author} {\bibfnamefont {S.}~\bibnamefont {Sheldon}}, \bibinfo
  {author} {\bibfnamefont {J.~M.}\ \bibnamefont {Chow}}, \ and\ \bibinfo
  {author} {\bibfnamefont {J.~M.}\ \bibnamefont {Gambetta}},\ }\href {\doibase
  10.1103/PhysRevA.96.022330} {\bibfield  {journal} {\bibinfo  {journal} {Phys.
  Rev. A}\ }\textbf {\bibinfo {volume} {96}},\ \bibinfo {pages} {022330}
  (\bibinfo {year} {2017})}\BibitemShut {NoStop}%
\bibitem [{\citenamefont {Loss}\ and\ \citenamefont
  {DiVincenzo}(1998)}]{Loss1998}%
  \BibitemOpen
  \bibfield  {author} {\bibinfo {author} {\bibfnamefont {D.}~\bibnamefont
  {Loss}}\ and\ \bibinfo {author} {\bibfnamefont {D.~P.}\ \bibnamefont
  {DiVincenzo}},\ }\href {\doibase 10.1103/PhysRevA.57.120} {\bibfield
  {journal} {\bibinfo  {journal} {Phys. Rev. A}\ }\textbf {\bibinfo {volume}
  {57}},\ \bibinfo {pages} {120} (\bibinfo {year} {1998})}\BibitemShut
  {NoStop}%
\bibitem [{\citenamefont {Watson}\ \emph {et~al.}(2018)\citenamefont {Watson},
  \citenamefont {Philips}, \citenamefont {Kawakami}, \citenamefont {Ward},
  \citenamefont {Scarlino}, \citenamefont {Veldhorst}, \citenamefont {Savage},
  \citenamefont {Lagally}, \citenamefont {Friesen}, \citenamefont
  {Coppersmith}, \citenamefont {Eriksson},\ and\ \citenamefont
  {Vandersypen}}]{Watson2018}%
  \BibitemOpen
  \bibfield  {author} {\bibinfo {author} {\bibfnamefont {T.~F.}\ \bibnamefont
  {Watson}}, \bibinfo {author} {\bibfnamefont {S.~G.}\ \bibnamefont {Philips}},
  \bibinfo {author} {\bibfnamefont {E.}~\bibnamefont {Kawakami}}, \bibinfo
  {author} {\bibfnamefont {D.~R.}\ \bibnamefont {Ward}}, \bibinfo {author}
  {\bibfnamefont {P.}~\bibnamefont {Scarlino}}, \bibinfo {author}
  {\bibfnamefont {M.}~\bibnamefont {Veldhorst}}, \bibinfo {author}
  {\bibfnamefont {D.~E.}\ \bibnamefont {Savage}}, \bibinfo {author}
  {\bibfnamefont {M.~G.}\ \bibnamefont {Lagally}}, \bibinfo {author}
  {\bibfnamefont {M.}~\bibnamefont {Friesen}}, \bibinfo {author} {\bibfnamefont
  {S.~N.}\ \bibnamefont {Coppersmith}}, \bibinfo {author} {\bibfnamefont
  {M.~A.}\ \bibnamefont {Eriksson}}, \ and\ \bibinfo {author} {\bibfnamefont
  {L.~M.}\ \bibnamefont {Vandersypen}},\ }\href {\doibase 10.1038/nature25766}
  {\bibfield  {journal} {\bibinfo  {journal} {Nature}\ }\textbf {\bibinfo
  {volume} {555}},\ \bibinfo {pages} {633} (\bibinfo {year} {2018})},\ \Eprint
  {http://arxiv.org/abs/1708.04214} {1708.04214} \BibitemShut {NoStop}%
\bibitem [{\citenamefont {Croot}\ \emph {et~al.}(2020)\citenamefont {Croot},
  \citenamefont {Mi}, \citenamefont {Putz}, \citenamefont {Benito},
  \citenamefont {Borjans}, \citenamefont {Burkard},\ and\ \citenamefont
  {Petta}}]{Croot2020}%
  \BibitemOpen
  \bibfield  {author} {\bibinfo {author} {\bibfnamefont {X.}~\bibnamefont
  {Croot}}, \bibinfo {author} {\bibfnamefont {X.}~\bibnamefont {Mi}}, \bibinfo
  {author} {\bibfnamefont {S.}~\bibnamefont {Putz}}, \bibinfo {author}
  {\bibfnamefont {M.}~\bibnamefont {Benito}}, \bibinfo {author} {\bibfnamefont
  {F.}~\bibnamefont {Borjans}}, \bibinfo {author} {\bibfnamefont
  {G.}~\bibnamefont {Burkard}}, \ and\ \bibinfo {author} {\bibfnamefont
  {J.~R.}\ \bibnamefont {Petta}},\ }\href {\doibase
  10.1103/PhysRevResearch.2.012006} {\bibfield  {journal} {\bibinfo  {journal}
  {Phys. Rev. Research}\ }\textbf {\bibinfo {volume} {2}},\ \bibinfo {pages}
  {012006} (\bibinfo {year} {2020})}\BibitemShut {NoStop}%
\bibitem [{\citenamefont {Undseth}\ \emph {et~al.}(2023)\citenamefont
  {Undseth}, \citenamefont {Xue}, \citenamefont {Mehmandoost}, \citenamefont
  {Russ}, \citenamefont {Eendebak}, \citenamefont {Samkharadze}, \citenamefont
  {Sammak}, \citenamefont {Dobrovitski}, \citenamefont {Scappucci},\ and\
  \citenamefont {Vandersypen}}]{Undseth2022}%
  \BibitemOpen
  \bibfield  {author} {\bibinfo {author} {\bibfnamefont {B.}~\bibnamefont
  {Undseth}}, \bibinfo {author} {\bibfnamefont {X.}~\bibnamefont {Xue}},
  \bibinfo {author} {\bibfnamefont {M.}~\bibnamefont {Mehmandoost}}, \bibinfo
  {author} {\bibfnamefont {M.}~\bibnamefont {Russ}}, \bibinfo {author}
  {\bibfnamefont {P.~T.}\ \bibnamefont {Eendebak}}, \bibinfo {author}
  {\bibfnamefont {N.}~\bibnamefont {Samkharadze}}, \bibinfo {author}
  {\bibfnamefont {A.}~\bibnamefont {Sammak}}, \bibinfo {author} {\bibfnamefont
  {V.~V.}\ \bibnamefont {Dobrovitski}}, \bibinfo {author} {\bibfnamefont
  {G.}~\bibnamefont {Scappucci}}, \ and\ \bibinfo {author} {\bibfnamefont
  {L.~M.~K.}\ \bibnamefont {Vandersypen}},\ }\href@noop {} {\  (\bibinfo {year}
  {2023})},\ \Eprint {http://arxiv.org/abs/2205.04905} {arXiv:2205.04905
  [cond-mat.mes-hall]} \BibitemShut {NoStop}%
\bibitem [{\citenamefont {Gilbert}\ \emph {et~al.}(2023)\citenamefont
  {Gilbert}, \citenamefont {Tanttu}, \citenamefont {Lim}, \citenamefont {Feng},
  \citenamefont {Huang}, \citenamefont {Cifuentes}, \citenamefont {Serrano},
  \citenamefont {Mai}, \citenamefont {Leon}, \citenamefont {Escott},
  \citenamefont {Itoh}, \citenamefont {Abrosimov}, \citenamefont {Pohl},
  \citenamefont {Thewalt}, \citenamefont {Hudson}, \citenamefont {Morello},
  \citenamefont {Laucht}, \citenamefont {Yang}, \citenamefont {Saraiva},\ and\
  \citenamefont {Dzurak}}]{Gilbert2022}%
  \BibitemOpen
  \bibfield  {author} {\bibinfo {author} {\bibfnamefont {W.}~\bibnamefont
  {Gilbert}}, \bibinfo {author} {\bibfnamefont {T.}~\bibnamefont {Tanttu}},
  \bibinfo {author} {\bibfnamefont {W.~H.}\ \bibnamefont {Lim}}, \bibinfo
  {author} {\bibfnamefont {M.}~\bibnamefont {Feng}}, \bibinfo {author}
  {\bibfnamefont {J.~Y.}\ \bibnamefont {Huang}}, \bibinfo {author}
  {\bibfnamefont {J.~D.}\ \bibnamefont {Cifuentes}}, \bibinfo {author}
  {\bibfnamefont {S.}~\bibnamefont {Serrano}}, \bibinfo {author} {\bibfnamefont
  {P.~Y.}\ \bibnamefont {Mai}}, \bibinfo {author} {\bibfnamefont {R.~C.~C.}\
  \bibnamefont {Leon}}, \bibinfo {author} {\bibfnamefont {C.~C.}\ \bibnamefont
  {Escott}}, \bibinfo {author} {\bibfnamefont {K.~M.}\ \bibnamefont {Itoh}},
  \bibinfo {author} {\bibfnamefont {N.~V.}\ \bibnamefont {Abrosimov}}, \bibinfo
  {author} {\bibfnamefont {H.-J.}\ \bibnamefont {Pohl}}, \bibinfo {author}
  {\bibfnamefont {M.~L.~W.}\ \bibnamefont {Thewalt}}, \bibinfo {author}
  {\bibfnamefont {F.~E.}\ \bibnamefont {Hudson}}, \bibinfo {author}
  {\bibfnamefont {A.}~\bibnamefont {Morello}}, \bibinfo {author} {\bibfnamefont
  {A.}~\bibnamefont {Laucht}}, \bibinfo {author} {\bibfnamefont {C.~H.}\
  \bibnamefont {Yang}}, \bibinfo {author} {\bibfnamefont {A.}~\bibnamefont
  {Saraiva}}, \ and\ \bibinfo {author} {\bibfnamefont {A.~S.}\ \bibnamefont
  {Dzurak}},\ }\href {\doibase 10.1038/s41565-022-01280-4} {\bibfield
  {journal} {\bibinfo  {journal} {Nat. Nanotechnol.}\ }\textbf {\bibinfo
  {volume} {18}},\ \bibinfo {pages} {131} (\bibinfo {year} {2023})}\BibitemShut
  {NoStop}%
\bibitem [{\citenamefont {Rackauckas}\ \emph {et~al.}(2019)\citenamefont
  {Rackauckas}, \citenamefont {Innes}, \citenamefont {Ma}, \citenamefont
  {Bettencourt}, \citenamefont {White},\ and\ \citenamefont
  {Dixit}}]{rackauckas2019diffeqfluxjl}%
  \BibitemOpen
  \bibfield  {author} {\bibinfo {author} {\bibfnamefont {C.}~\bibnamefont
  {Rackauckas}}, \bibinfo {author} {\bibfnamefont {M.}~\bibnamefont {Innes}},
  \bibinfo {author} {\bibfnamefont {Y.}~\bibnamefont {Ma}}, \bibinfo {author}
  {\bibfnamefont {J.}~\bibnamefont {Bettencourt}}, \bibinfo {author}
  {\bibfnamefont {L.}~\bibnamefont {White}}, \ and\ \bibinfo {author}
  {\bibfnamefont {V.}~\bibnamefont {Dixit}},\ }\href@noop {} {\  (\bibinfo
  {year} {2019})},\ \Eprint {http://arxiv.org/abs/1902.02376} {arXiv:1902.02376
  [cs.LG]} \BibitemShut {NoStop}%
\bibitem [{\citenamefont {Rackauckas}\ and\ \citenamefont
  {Nie}(2017)}]{Rackauckas2017}%
  \BibitemOpen
  \bibfield  {author} {\bibinfo {author} {\bibfnamefont {C.}~\bibnamefont
  {Rackauckas}}\ and\ \bibinfo {author} {\bibfnamefont {Q.}~\bibnamefont
  {Nie}},\ }\href {\doibase 10.5334/jors.151} {\bibfield  {journal} {\bibinfo
  {journal} {The Journal of Open Research Software}\ }\textbf {\bibinfo
  {volume} {5}},\ \bibinfo {pages} {15} (\bibinfo {year} {2017})}\BibitemShut
  {NoStop}%
\bibitem [{\citenamefont {Boter}\ \emph {et~al.}(2020)\citenamefont {Boter},
  \citenamefont {Xue}, \citenamefont {Kr\"ahenmann}, \citenamefont {Watson},
  \citenamefont {Premakumar}, \citenamefont {Ward}, \citenamefont {Savage},
  \citenamefont {Lagally}, \citenamefont {Friesen}, \citenamefont
  {Coppersmith}, \citenamefont {Eriksson}, \citenamefont {Joynt},\ and\
  \citenamefont {Vandersypen}}]{Jelmer2020}%
  \BibitemOpen
  \bibfield  {author} {\bibinfo {author} {\bibfnamefont {J.~M.}\ \bibnamefont
  {Boter}}, \bibinfo {author} {\bibfnamefont {X.}~\bibnamefont {Xue}}, \bibinfo
  {author} {\bibfnamefont {T.}~\bibnamefont {Kr\"ahenmann}}, \bibinfo {author}
  {\bibfnamefont {T.~F.}\ \bibnamefont {Watson}}, \bibinfo {author}
  {\bibfnamefont {V.~N.}\ \bibnamefont {Premakumar}}, \bibinfo {author}
  {\bibfnamefont {D.~R.}\ \bibnamefont {Ward}}, \bibinfo {author}
  {\bibfnamefont {D.~E.}\ \bibnamefont {Savage}}, \bibinfo {author}
  {\bibfnamefont {M.~G.}\ \bibnamefont {Lagally}}, \bibinfo {author}
  {\bibfnamefont {M.}~\bibnamefont {Friesen}}, \bibinfo {author} {\bibfnamefont
  {S.~N.}\ \bibnamefont {Coppersmith}}, \bibinfo {author} {\bibfnamefont
  {M.~A.}\ \bibnamefont {Eriksson}}, \bibinfo {author} {\bibfnamefont
  {R.}~\bibnamefont {Joynt}}, \ and\ \bibinfo {author} {\bibfnamefont
  {L.~M.~K.}\ \bibnamefont {Vandersypen}},\ }\href {\doibase
  10.1103/PhysRevB.101.235133} {\bibfield  {journal} {\bibinfo  {journal}
  {Phys. Rev. B}\ }\textbf {\bibinfo {volume} {101}},\ \bibinfo {pages}
  {235133} (\bibinfo {year} {2020})}\BibitemShut {NoStop}%
\bibitem [{\citenamefont {Meunier}\ \emph {et~al.}(2011)\citenamefont
  {Meunier}, \citenamefont {Calado},\ and\ \citenamefont
  {Vandersypen}}]{meunier_efficient_2011}%
  \BibitemOpen
  \bibfield  {author} {\bibinfo {author} {\bibfnamefont {T.}~\bibnamefont
  {Meunier}}, \bibinfo {author} {\bibfnamefont {V.~E.}\ \bibnamefont {Calado}},
  \ and\ \bibinfo {author} {\bibfnamefont {L.~M.~K.}\ \bibnamefont
  {Vandersypen}},\ }\href {\doibase 10.1103/PhysRevB.83.121403} {\bibfield
  {journal} {\bibinfo  {journal} {Phys. Rev. B}\ }\textbf {\bibinfo {volume}
  {83}},\ \bibinfo {pages} {121403} (\bibinfo {year} {2011})}\BibitemShut
  {NoStop}%
\bibitem [{\citenamefont {Green}\ \emph {et~al.}(2013)\citenamefont {Green},
  \citenamefont {Sastrawan}, \citenamefont {Uys},\ and\ \citenamefont
  {Biercuk}}]{Green_2013}%
  \BibitemOpen
  \bibfield  {author} {\bibinfo {author} {\bibfnamefont {T.~J.}\ \bibnamefont
  {Green}}, \bibinfo {author} {\bibfnamefont {J.}~\bibnamefont {Sastrawan}},
  \bibinfo {author} {\bibfnamefont {H.}~\bibnamefont {Uys}}, \ and\ \bibinfo
  {author} {\bibfnamefont {M.~J.}\ \bibnamefont {Biercuk}},\ }\href {\doibase
  10.1088/1367-2630/15/9/095004} {\bibfield  {journal} {\bibinfo  {journal}
  {New J. Phys.}\ }\textbf {\bibinfo {volume} {15}},\ \bibinfo {pages} {095004}
  (\bibinfo {year} {2013})}\BibitemShut {NoStop}%
\bibitem [{\citenamefont {Connors}\ \emph {et~al.}(2019)\citenamefont
  {Connors}, \citenamefont {Nelson}, \citenamefont {Qiao}, \citenamefont
  {Edge},\ and\ \citenamefont {Nichol}}]{connors2020chargenoise}%
  \BibitemOpen
  \bibfield  {author} {\bibinfo {author} {\bibfnamefont {E.~J.}\ \bibnamefont
  {Connors}}, \bibinfo {author} {\bibfnamefont {J.}~\bibnamefont {Nelson}},
  \bibinfo {author} {\bibfnamefont {H.}~\bibnamefont {Qiao}}, \bibinfo {author}
  {\bibfnamefont {L.~F.}\ \bibnamefont {Edge}}, \ and\ \bibinfo {author}
  {\bibfnamefont {J.~M.}\ \bibnamefont {Nichol}},\ }\href {\doibase
  10.1103/PhysRevB.100.165305} {\bibfield  {journal} {\bibinfo  {journal}
  {Phys. Rev. B}\ }\textbf {\bibinfo {volume} {100}},\ \bibinfo {pages}
  {165305} (\bibinfo {year} {2019})}\BibitemShut {NoStop}%
\bibitem [{\citenamefont {Paquelet~Wuetz}\ \emph {et~al.}(2023)\citenamefont
  {Paquelet~Wuetz}, \citenamefont {Degli~Esposti}, \citenamefont {Zwerver},
  \citenamefont {Amitonov}, \citenamefont {Botifoll}, \citenamefont {Arbiol},
  \citenamefont {Vandersypen}, \citenamefont {Russ},\ and\ \citenamefont
  {Scappucci}}]{paquelet_wuetz_reducing_2023}%
  \BibitemOpen
  \bibfield  {author} {\bibinfo {author} {\bibfnamefont {B.}~\bibnamefont
  {Paquelet~Wuetz}}, \bibinfo {author} {\bibfnamefont {D.}~\bibnamefont
  {Degli~Esposti}}, \bibinfo {author} {\bibfnamefont {A.-M.~J.}\ \bibnamefont
  {Zwerver}}, \bibinfo {author} {\bibfnamefont {S.~V.}\ \bibnamefont
  {Amitonov}}, \bibinfo {author} {\bibfnamefont {M.}~\bibnamefont {Botifoll}},
  \bibinfo {author} {\bibfnamefont {J.}~\bibnamefont {Arbiol}}, \bibinfo
  {author} {\bibfnamefont {L.~M.~K.}\ \bibnamefont {Vandersypen}}, \bibinfo
  {author} {\bibfnamefont {M.}~\bibnamefont {Russ}}, \ and\ \bibinfo {author}
  {\bibfnamefont {G.}~\bibnamefont {Scappucci}},\ }\href {\doibase
  10.1038/s41467-023-36951-w} {\bibfield  {journal} {\bibinfo  {journal} {Nat.
  Commun.}\ }\textbf {\bibinfo {volume} {14}},\ \bibinfo {pages} {1385}
  (\bibinfo {year} {2023})}\BibitemShut {NoStop}%
\bibitem [{\citenamefont {Xue}\ \emph {et~al.}(2021)\citenamefont {Xue},
  \citenamefont {Patra}, \citenamefont {van Dijk}, \citenamefont {Samkharadze},
  \citenamefont {Subramanian}, \citenamefont {Corna}, \citenamefont
  {Paquelet~Wuetz}, \citenamefont {Jeon}, \citenamefont {Sheikh}, \citenamefont
  {Juarez-Hernandez}, \citenamefont {Esparza}, \citenamefont {Rampurawala},
  \citenamefont {Carlton}, \citenamefont {Ravikumar}, \citenamefont {Nieva},
  \citenamefont {Kim}, \citenamefont {Lee}, \citenamefont {Sammak},
  \citenamefont {Scappucci}, \citenamefont {Veldhorst}, \citenamefont
  {Sebastiano}, \citenamefont {Babaie}, \citenamefont {Pellerano},
  \citenamefont {Charbon},\ and\ \citenamefont
  {Vandersypen}}]{xue_cmos-based_2021}%
  \BibitemOpen
  \bibfield  {author} {\bibinfo {author} {\bibfnamefont {X.}~\bibnamefont
  {Xue}}, \bibinfo {author} {\bibfnamefont {B.}~\bibnamefont {Patra}}, \bibinfo
  {author} {\bibfnamefont {J.~P.~G.}\ \bibnamefont {van Dijk}}, \bibinfo
  {author} {\bibfnamefont {N.}~\bibnamefont {Samkharadze}}, \bibinfo {author}
  {\bibfnamefont {S.}~\bibnamefont {Subramanian}}, \bibinfo {author}
  {\bibfnamefont {A.}~\bibnamefont {Corna}}, \bibinfo {author} {\bibfnamefont
  {B.}~\bibnamefont {Paquelet~Wuetz}}, \bibinfo {author} {\bibfnamefont
  {C.}~\bibnamefont {Jeon}}, \bibinfo {author} {\bibfnamefont {F.}~\bibnamefont
  {Sheikh}}, \bibinfo {author} {\bibfnamefont {E.}~\bibnamefont
  {Juarez-Hernandez}}, \bibinfo {author} {\bibfnamefont {B.~P.}\ \bibnamefont
  {Esparza}}, \bibinfo {author} {\bibfnamefont {H.}~\bibnamefont
  {Rampurawala}}, \bibinfo {author} {\bibfnamefont {B.}~\bibnamefont
  {Carlton}}, \bibinfo {author} {\bibfnamefont {S.}~\bibnamefont {Ravikumar}},
  \bibinfo {author} {\bibfnamefont {C.}~\bibnamefont {Nieva}}, \bibinfo
  {author} {\bibfnamefont {S.}~\bibnamefont {Kim}}, \bibinfo {author}
  {\bibfnamefont {H.-J.}\ \bibnamefont {Lee}}, \bibinfo {author} {\bibfnamefont
  {A.}~\bibnamefont {Sammak}}, \bibinfo {author} {\bibfnamefont
  {G.}~\bibnamefont {Scappucci}}, \bibinfo {author} {\bibfnamefont
  {M.}~\bibnamefont {Veldhorst}}, \bibinfo {author} {\bibfnamefont
  {F.}~\bibnamefont {Sebastiano}}, \bibinfo {author} {\bibfnamefont
  {M.}~\bibnamefont {Babaie}}, \bibinfo {author} {\bibfnamefont
  {S.}~\bibnamefont {Pellerano}}, \bibinfo {author} {\bibfnamefont
  {E.}~\bibnamefont {Charbon}}, \ and\ \bibinfo {author} {\bibfnamefont
  {L.~M.~K.}\ \bibnamefont {Vandersypen}},\ }\href {\doibase
  10.1038/s41586-021-03469-4} {\bibfield  {journal} {\bibinfo  {journal}
  {Nature}\ }\textbf {\bibinfo {volume} {593}},\ \bibinfo {pages} {205}
  (\bibinfo {year} {2021})}\BibitemShut {NoStop}%
\end{thebibliography}%
\end{document}